\documentclass[%
 aip,
 amsmath,amssymb,
 reprint,%
]{revtex4-1}

\usepackage{graphicx}
\usepackage{epstopdf}
\usepackage{dcolumn}
\usepackage{bm}
\usepackage{float}

\usepackage[utf8]{inputenc}
\usepackage[T1]{fontenc}
\usepackage{mathptmx}
\usepackage{etoolbox}
\usepackage{epstopdf}
\usepackage{array}
\usepackage{booktabs}
\usepackage{lipsum}
\usepackage{float}
\usepackage{xcolor}
\usepackage{makecell}
\usepackage{siunitx} 
\usepackage{dsfont}
\usepackage{CJKutf8}

\makeatletter
\def\@email#1#2{%
 \endgroup
 \patchcmd{\titleblock@produce}
  {\frontmatter@RRAPformat}
  {\frontmatter@RRAPformat{\produce@RRAP{*#1\href{mailto:#2}{#2}}}\frontmatter@RRAPformat}
  {}{}
}%
\makeatother

\usepackage{xcolor} 

\begin{document}

\preprint{AIP/123-QED}

\title{Machine-learned flow estimation with sparse data --- exemplified for the rooftop of an unmanned aerial vehicle vertiport}

\begin{CJK*}{UTF8}{gbsn}
\author{Chang Hou (侯昶)}
\affiliation{Chair of Artificial Intelligence and Aerodynamics, School of Mechanical Engineering and Automation, Harbin Institute of Technology, Shenzhen 518055, Peoples' Republic of China}

\author{Luigi Marra}
\affiliation{Department of Aerospace Engineering, Universidad Carlos III de Madrid, Av. de la Universidad, 30, Legan{é}s, 28911, Madrid, Spain}
 
\author{Guy Y. Cornejo Maceda}
\affiliation{Chair of Artificial Intelligence and Aerodynamics, School of Mechanical Engineering and Automation, Harbin Institute of Technology, Shenzhen 518055, Peoples' Republic of China}
\affiliation{Department of Aerospace Engineering, Universidad Carlos III de Madrid, Av. de la Universidad, 30, Legan{é}s, 28911, Madrid, Spain}
\email{yoslan@hit.edu.cn}

\author{Peng Jiang (姜鹏)}
\affiliation{Chair of Artificial Intelligence and Aerodynamics, School of Mechanical Engineering and Automation, Harbin Institute of Technology, Shenzhen 518055, Peoples' Republic of China}

\author{Jingguo Chen (陈靖国)}
\affiliation{Chair of Artificial Intelligence and Aerodynamics, School of Mechanical Engineering and Automation, Harbin Institute of Technology, Shenzhen 518055, Peoples' Republic of China}

\author{Yutong Liu (刘宇同)}
\affiliation{Chair of Artificial Intelligence and Aerodynamics, School of Mechanical Engineering and Automation, Harbin Institute of Technology, Shenzhen 518055, Peoples' Republic of China}

\author{Gang Hu (胡钢)}
\affiliation{Artificial Intelligence for Wind Engineering (AIWE) Lab, School of Civil and Environmental Engineering, Harbin Institute of Technology, Shenzhen 518055, Peoples' Republic of China}
\affiliation{Guangdong Provincial Key Laboratory of Intelligent and Resilient Structures for Civil Engineering, Harbin Institute of Technology, Shenzhen 518055, Peoples' Republic of China}
\author{Jialong Chen (陈佳龙)}
\affiliation{Meituan Technology Co., Ltd, Shenzhen 518131, People’s Republic of China}

\author{Andrea Ianiro}
\affiliation{Department of Aerospace Engineering, Universidad Carlos III de Madrid, Av. de la Universidad, 30, Legan{é}s, 28911, Madrid, Spain}

\author{Stefano Discetti}
\affiliation{Department of Aerospace Engineering, Universidad Carlos III de Madrid, Av. de la Universidad, 30, Legan{é}s, 28911, Madrid, Spain}

\author{Andrea Meil{á}n-Vila}
\affiliation{Department of Statistics, Universidad Carlos III de Madrid, Av. de la Universidad, 30, Legan{é}s, 28911, Madrid, Spain}

\author{Bernd R. Noack}%
\affiliation{Chair of Artificial Intelligence and Aerodynamics, School of Mechanical Engineering and Automation, Harbin Institute of Technology, Shenzhen 518055, Peoples' Republic of China}
\affiliation{Guangdong Provincial Key Laboratory of Intelligent Morphing Mechanisms and Adaptive Robotics, Harbin Institute of Technology, 518055 Shenzhen, Peoples' Republic of China}
\email{bernd.noack@hit.edu.cn}
\affiliation{Department of Aerospace Engineering, Universidad Carlos III de Madrid, Av. de la Universidad, 30, Legan{é}s, 28911, Madrid, Spain}

\date{\today}

\begin{abstract}
We propose a physics-informed data-driven framework for urban wind estimation.
This framework validates and incorporates the Reynolds number independence for flows under various working conditions,
thus allowing the extrapolation for wind conditions far beyond the training data.
Another key enabler is a machine-learned non-dimensionalized manifold from snapshot data.
The velocity field is modeled using a double encoder-decoder approach.
The first encoder normalizes data using the oncoming wind speed, 
while the second encoder projects this normalized data onto the isometric feature mapping manifold. 
The decoders reverse this process, with $k$-nearest neighbor performing the first decoding and the second undoing the normalization.
The manifold is coarse-grained by clustering
to reduce the computational load for de- and encoding.
The sensor-based flow estimation 
is based on the estimate of the oncoming wind speed
and a mapping from sensor signal
to the manifold latent variables.
The proposed machine-learned flow estimation framework is exemplified for the flow above an Unmanned Aerial Vehicle vertiport. 
The wind estimation is shown to generalize well for rare wind conditions, not included in the original database.
\end{abstract}

\maketitle

\end{CJK*}
 
\section{Introduction}
\label{sec1}

\begin{figure}
\includegraphics[width=8.5cm]{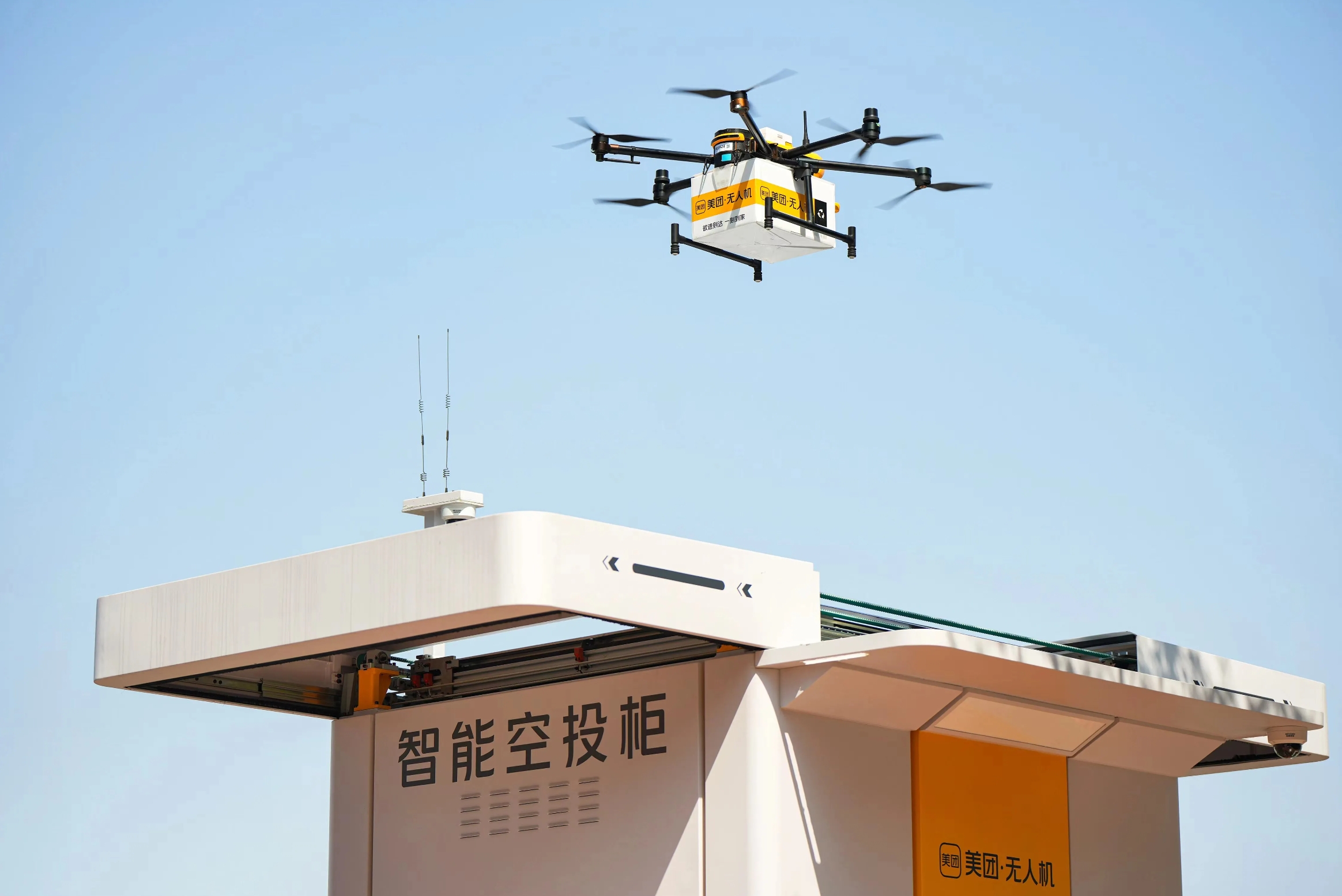}
\caption{UAV vertiport and drone delivery by Meituan.
Reproduced with permission from Meituan Technology Co. Ltd.
}
\label{Fig:MeituanVertiport}  
\end{figure}

In the near future, humanity will conquer three-dimensional mobility, transforming transportation in both urban and rural areas\citep{yang2024sizing,motlagh2016low}. 
The low-altitude economy, where drones and air taxis become a routine part of daily life, is on the horizon\citep{antcliff2016silicon}, as shown in figure~\ref{Fig:MeituanVertiport}. 
This emerging economy builds on the foundation of the traditional general aviation industry, integrating innovative production and service methods powered by drones and air taxis.
However, drones and air taxis are more vulnerable to strong winds and sudden wind gusts as compared to airplanes. 
This vulnerability increases the risk of accidents, especially near vertiports, buildings, or other airborne vehicles. 
The complex flow patterns in urban environments, characterized by sharp velocity gradients, such as recirculation zones over vertiports or wakes behind high-rise buildings, further complicate safe navigation\citep{balestrieri2021sensors,ambrogi2024influence,jones2022physics}. 
These rapid spatial and temporal changes in wind speed make it challenging to maintain drones on their intended low-altitude paths.
Therefore, developing urban wind estimation technology is essential for ensuring the safety and efficiency of this modern aerial society \citep{ou2024dynamic}.

\begin{table*}
\caption{Research on flow estimation of urban environments, for details see texts.}
\label{Table1}
\begingroup
\renewcommand{\arraystretch}{1.1} 
\begin{ruledtabular}
\begin{tabular}{lcccc}
\textbf{Team} & \textbf{Configuration}  & \textbf{Method} & \makecell{\textbf{Extrapolation} \\\textbf{beyond database}}\\[6pt]
\hline
\citet{toparlar2015cfd}           &  Bergpolder Zuid region in Rotterdam          & CFD (uRANS)              & \\[4pt] 
\citet{ramponi2015cfd}            &  Generic buildings with parallel streets          & CFD (RANS)               & \\[4pt] 
\citet{jiang2024identification}   &  Mong Kok district in Hong Kong          & CFD (RANS)               & \\[4pt] 
\citet{van2010effect}             &  A large semi-enclosed stadium          & CFD (RANS)               & \\[4pt] 
\citet{rovere2022safety}          &  Ground with a helicopter  & CFD (uRANS)              & \\[4pt] 
\citet{andronikos2021validation}  &  An obstacle with a helicopter & CFD                      & \\[4pt] 
\citet{yunus2023efficient}        &  Urban environment with vertiport & CFD (Gaussian beam tracing)      & \\[4pt] 
\citet{weerasuriya2018wind}       &  Tsuen Wan in Hong Kong          & Experimental (Wind tunnel)        & \\[4pt] 
\citet{ricci2017wind}             &  Livorno in Italy          & Experimental (Wind tunnel)        & \\[4pt] 
\citet{gao2012field}              &  Building complex          & Experimental (On-site sensor)           & \\[4pt] 
\citet{zou2021field}              &  Darlington campus of University of Sydney           & Experimental (On-site sensor)           & \\[4pt] 
\citet{shao2023pignn}             &  High-rise building and city blocks          & Data-centric (GNN)       & \\[4pt] 
\citet{gao2024urban}              &  Niigata in Japan          & Data-centric (GNN)       & \\[4pt] 
Current study                     &  Meituan UAV vertiport & Data-centric (Nondimensionalized manifold)     & \checkmark\\[4pt] 
\end{tabular}
\end{ruledtabular}
\endgroup
\end{table*}
%

Research on urban wind estimation primarily falls into three categories: computational fluid dynamics (CFD) simulations, experiments (wind tunnel testing, field measurements), and data-centric methods\citep{gao2023optimal}.
Table~\ref{Table1} summarizes representative research. 
CFD methods can provide full-state information under varying wind conditions, such as wind angle and velocity. 
For example, \citet{toparlar2015cfd} conducted unsteady Reynolds-averaged Navier–Stokes (uRANS) simulations on wind flow and temperatures in the Bergpolder Zuid region, demonstrating CFD's potential for accurately predicting urban microclimates. 
Similarly, \citet{ramponi2015cfd} used 3D steady RANS simulations on generic urban configurations, highlighting the effects of wind angle and urban morphology on wind field patterns. 
\citet{jiang2024identification} introduced a methodology for defining no-fly zones for drones using CFD, offering three hazardous indices—safe, deviation, and unsafe—to assess drone operation conditions. 
Additionally, \citet{van2010effect} used CFD to evaluate the influence of wind angle and urban geometry on air exchange rates within a stadium model.
Simplified generic geometries, often composed of regular arrays of obstacles, are also commonly used for systematic studies of aerodynamic and geometric correlations, as they are computationally cheaper \citep{buccolieri2010city,hang2010wind,moonen2011evaluation,lin2014quantitative}. 
Wind tunnel experiments and on-site measurements likewise provide valuable insights into urban wind patterns, with notable methods including wind tunnel and fan-array wind generator experiments\citep{weerasuriya2018wind,ricci2017wind,wang2024coarse,li2024aerodynamic,liu2024aerodynamic},  as well as the on-site wind data collected by sensors\citep{gao2012field,zou2021field}.

An extensive range of data-driven methods focusing on urban wind estimation is emerging. 
\citet{shao2023pignn} introduced a novel physics-informed graph neural network (GNN) that operates $1-2$ orders of magnitude faster than traditional CFD methods while maintaining a nearly consistent level of accuracy. 
Additionally, data-driven approaches can leverage sensor-based estimations\citep{li2022machine}, which have recently become a research hotspot in the flow estimation community. 
\citet{gao2024urban} presents a deep learning (DL) model specifically designed to reconstruct velocity fields under varying wind angles using sparse sensor signals. 
Nevertheless, high-fidelity CFD simulations and experiments are costly and complex due to the wide range of operational conditions. 
In contrast, data-driven methods offer more economical and practical solutions for urban wind estimation.

While drones and air taxis promise to revolutionize transportation, their safe and efficient operation in urban environments relies on ground infrastructure\citep{schweiger2022urban}. 
Unmanned Aerial Vehicle (UAV) vertiports act as critical nodes connecting aerial mobility. 
Unlike conventional helicopters, 
which have sparse designated locations for take-off and landing remote from people, UAV vertiports are typically placed on the ground or the rooftops of skyscrapers\citep{araghizadeh2024aeroacoustic}. 
Research on flow estimation of vertiports is limited, most focus on ground effects and noise mitigation.
During vertical maneuvers near a vertiport, strong interactions between the aircraft and the surface—referred to as ground effects—can significantly influence flight dynamics, aerodynamic performance, and wake development. 
\citet{rovere2022safety} conducted a uRANS evaluation to support vertiport safety analysis, examining drone operations under the ground effects in various flight conditions. 
Similarly, \citet{andronikos2021validation} used CFD to reproduce the load effects on a helicopter’s main rotor above a vertiport and characterized the flow field in the rotor’s wake as it interacts with the vertiport.
Implementing vertiports in densely populated residential areas also requires careful analysis of noise propagation. 
\citet{yunus2023efficient} introduced a computationally efficient methodology to assess the acoustic impact of design, operational, and environmental factors on the vertiport’s noise footprint, importing 3D wind profiles directly from high-fidelity CFD simulations.

The safe operation of drones and air taxis during take-off and landing relies on efficient flight control algorithms, which require accurate flow information on the rooftop of vertiports. 
While most current research emphasizes large-scale urban wind estimation for flight route planning and drone cruise, vertiport flow estimation
has been largely neglected,
although displacements by wind gusts may lead to accidents.
Therefore,  flow estimation above vertiports is a crucial yet underexplored area in urban wind studies.
Moreover, given the complexity of urban environments, neither numerical nor experimental methods, nor data-driven approaches, can extrapolate wind conditions far beyond existing databases. 
Thus, a more generalized flow estimation framework with robust extrapolation capabilities is required.

The key enabler is non-dimensionality. 
Once a critical Reynolds number is reached, properly-scaled flow features are Reynolds-number independent ($Re$-independence)~\citep{uehara2003studies}. 
Following~\citet{castro1977flow}, a typical critical Reynolds number of a $2000$~mm high cubic model building immersed in a fully developed boundary layer is about $4 \times 10^4$. 
The wind speed at the cube height corresponding to that Reynolds number is about $3$ m/s, which is typically smaller than the oncoming wind speed in urban environments.
Therefore, incorporating non-dimensionalization into the flow estimation framework may thus facilitate extrapolation to the wind conditions outside the training dataset. 
A key component is data-driven manifold modeling, a promising approach for reduced-order representations\citep{taira2017modal,sun2024generalized}. 
For instance, 
a two-dimensional manifold can effectively capture 
transient and post-transient vortex shedding behaviors, with a dimension that is a small fraction of Proper Orthogonal Decomposition (POD) modes required for similar resolution\cite{2009_JCP_Bergmann_enablers,Bourgeois2013jfm}. 
Manifolds can be constructed using mean-field approaches \citep{noack2003hierarchy,bobusch2013experimental,protas2015optimal,deng2020low,he2021data}, cluster-based representations \citep{kaiser2014cluster,nair2019cluster,li2021cluster,hou2024dynamics}, dynamic features \citep{laval2001nonlocality,bagheri2013koopman,kindree2018low,loiseau2018sparse}, 
convex displacement interpolation \citep{cucchiara2024model},
 isometric feature mapping \citep{farzamnik2023snapshots,li2024manifold,ma2024efficient}
 and even parametric manifolds \cite{marra2024actuation,fukami2023grasping}.
These manifolds are robust to parameter changes, and the latent variables often have a clear physical meaning, addressing the issue of limited physical interpretability in traditional data-driven methods. 
Therefore, developing a manifold-based approach for flow estimation across a broad range of operating conditions 
holds promise and is the focus of the current study.

\begin{figure}
\includegraphics[width=8.5cm]{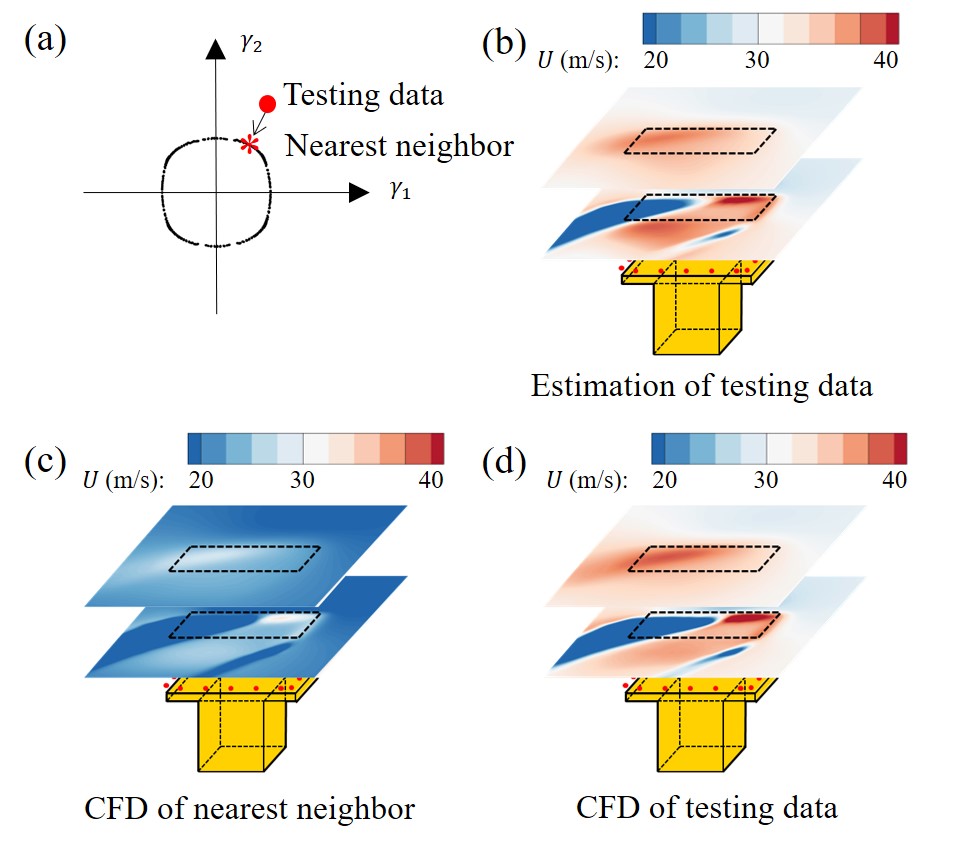}
\caption{An example of flow estimation with extrapolation, exemplified for the rooftop of UAV vertiport. 
(a) Testing data (red dot), the nondimensionalized database (black dots) and the nearest neighbor of the testing data (red star) displayed in a 2D representation of the manifold.  
(b) Estimated flow field.  
(c) Velocity field of the nearest neighbor.  
(d) Velocity field of the testing data. 
For testing sensor signals outside the existing database, the nearest neighbor may yield inaccurate flow information. 
Instead, we use the estimated oncoming wind speed to non-dimensionalize the testing data and subsequently restore the estimated flow field. 
By incorporating Reynolds number independence, this approach effectively generalizes to rare wind conditions.
}
\label{Fig:introdution}  
\end{figure}

In this manuscript, we propose a machine-learned flow estimation framework exemplified for a UAV vertiport, leveraging non-dimensionalization using the oncoming wind speed $U_\infty$ and a cluster-based manifold learner. 
The proposed framework maps non-dimensionalized data into a low-dimensional space through the isometric feature mapping $k$-nearest-neighbors (ISOMAP-$k$NN) method \citep{farzamnik2023snapshots}. 
Sensor signals are employed to estimate the wind conditions, enabling the subsequent estimation of the manifold's latent variables. 
The ISOMAP-$k$NN decoder further reconstructs the full-state, high-dimensional flow field data from the estimated latent variables.
The framework cooperates with Reynolds number independence, allowing extrapolations of wind conditions not included in the training data. This is illustrated in Figure~\ref{Fig:introdution}, where the full-state flow estimation above a UAV vertiport with rare wind conditions is estimated from sparse data (sensor signals), after non-dimensionalization. 
Additionally, clustering is incorporated to create a coarse-grained manifold, significantly reducing computational costs. 
The adaptability of this framework to a wide range of wind conditions enhances its applicability in real-world scenarios.

The manuscript is structured as follows.
In $\S$ 2, the UAV vertiport configurations and the corresponding database are described.
In $\S$ 3, we outline the methodology employed for the machine-learned flow estimation framework.
The results are presented and discussed in $\S$ 4, followed by the conclusion and potential future directions in $\S$ 5.
%

\section{Configuration}
\label{sec2}

This section provides an overview of the flow around a vertiport. 
The simulation platform is described in \S \ref{sec2.1}, followed by a discussion on the validation of the numerical method in \S \ref{sec2.2}.
Finally, the representative flow characteristics are presented in \S \ref{sec2.3}.
The variables used in this manuscript are explained in table~\ref{Table2}.
\begin{table}
\caption{Table of variables.}
\label{Table2}
\begingroup
\renewcommand{\arraystretch}{1.1} 
\begin{ruledtabular}
\begin{tabular}{cl} 
\textbf{Variable} & \textbf{Description} \\[6pt]
\hline
\multicolumn{2}{l}{ ----------------------------- Physical variables ----------------------------- }\\   
$U_{\infty}$  & Wind speed  \\  
$\alpha$  & Wind angle  \\  
$b$  & Wind level (Beaufort Scale)  \\
$B$  & Total number of wind levels  \\
$H$  & Characteristic length (The height of the vertiport)  \\  
$\bm{u}$  & Velocity vector  \\  
$p$  & Static pressure  \\  
$M$  & Number of snapshots  \\  
$\bm{u}^{m}$  & $m$-th snapshots of training velocity field  \\  
$\bm{s}^{m}$  & $m$-th training sensor signal  \\  
$\mathcal{C}_k$  & $k$-th cluster  \\ 
$\bm{c}_{k}$  & $k$-th centroid  \\  
$\bm{s}(t)$  & $t$-th testing sensor signal  \\  
$\bm{n}(t)$  & $t$-th noise signal  \\ 
$\bm{\hat{u}}(t)$  & $t$-th estimated velocity field \\  
$\Omega$  & Domain \\  
$K_c$ & Number of clusters  \\  
$K_e$  & Number of nearest neighbors in ISOMAP encoder  \\  
$K_d$  & Number of nearest neighbors in ISOMAP decoder  \\ 
$K_p$  & Number of nearest neighbors for the snapshots manifolds  \\ 
$K_w$  & Number of nearest neighbors for flow estimation  \\ 
$\chi$ & Characteristic function \\ 
$\mathcal{N}$  & Set of nearest neighbors \\  
$\lambda^l$  & Scaling factor in wind condition estimator  \\  
$\hat{U}_{\infty}$  & Estimated wind speed  \\  
$\hat{\alpha}$  & Estimated wind angle  \\  
$\mathbf{D}$  & Distance matrix  \\  
$\bm{\gamma}^m$  & Latent variables of the manifold \\  
$\bm{\bar{\gamma}}_k$  & Latent variables of the manifold discretized by the centroids \\  
$E$  & Representation error \\  
\multicolumn{2}{l}{ --------------------- Non-dimensionalized variables --------------------- }\\   
$\bm{u}^{*m}$  & $m$-th snapshot of training velocity field  \\  
$\bm{s}^{*m}$  & $m$-th training sensor signal  \\  
$\mathcal{C}^*_k$  & $k$-th cluster  \\  
$\bm{c}^*_{k}$  & $k$-th centroid  \\  
$\bm{s}^*(t)$  &  $t$-th testing sensor signal  \\  
$\bm{\hat{u}}^*(t)$ & $t$-th estimated velocity field \\  
\end{tabular}
\end{ruledtabular}
\endgroup
\end{table}
%

\subsection{Numerical simulation}
\label{sec2.1}

\begin{figure}[h]
\includegraphics[width=8.5cm]{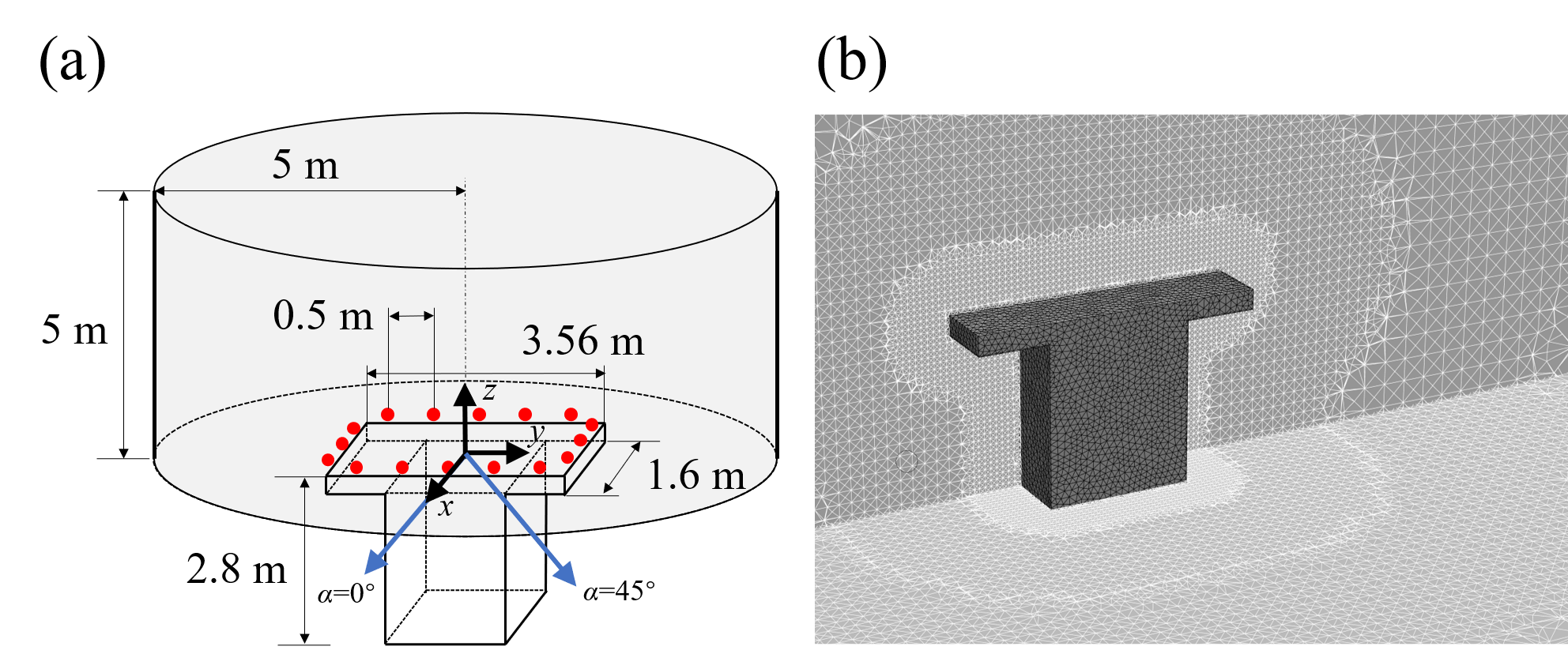}
\includegraphics[width=8.5cm]{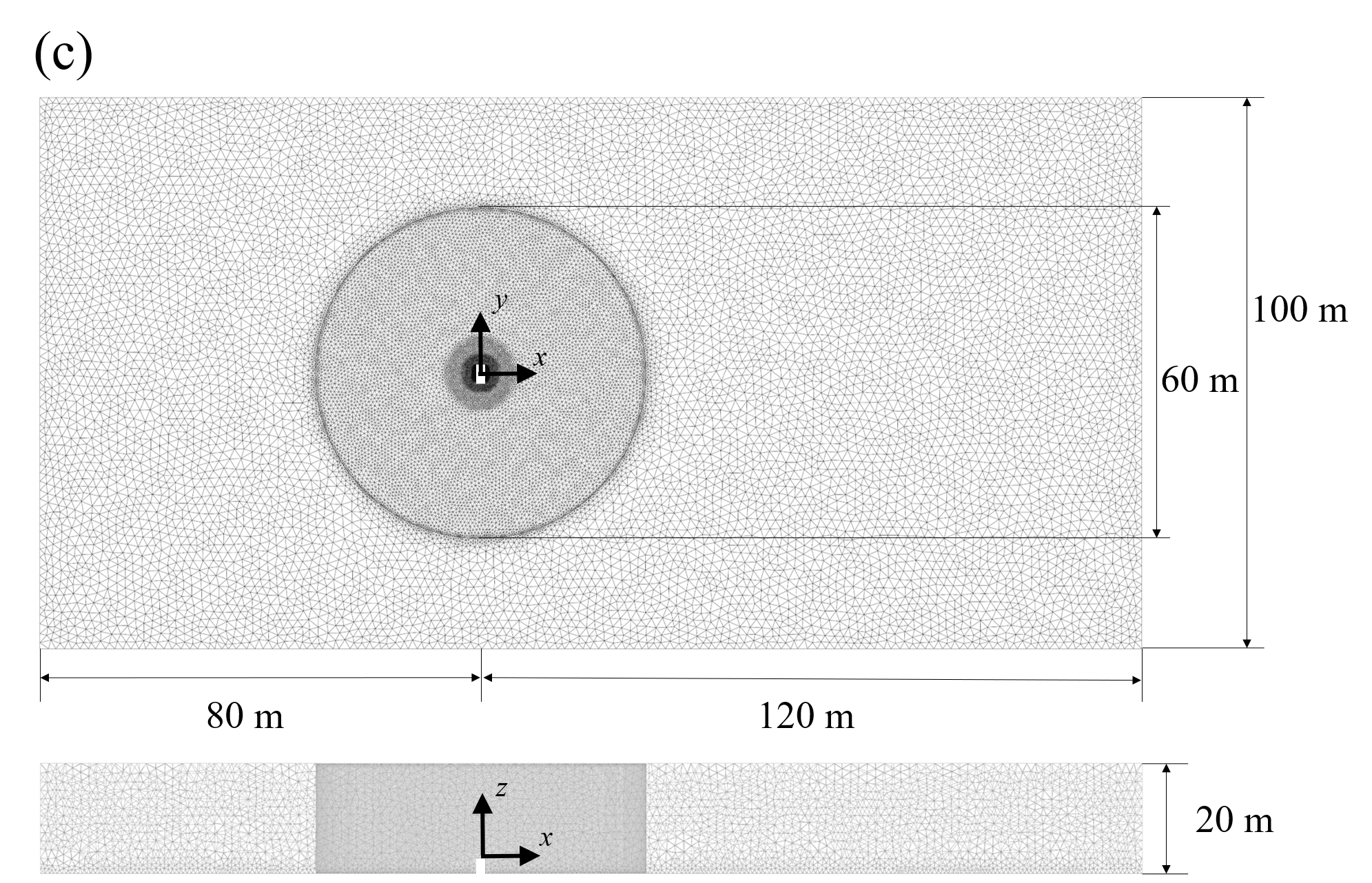}
\caption{Sketch of the UAV vertiport.
(a) Study domain. The velocity field data is obtained from the cylinder region.
The red dots represent the sensor location.
(b) Computational grid around the vertiport in a transverse and the coronal plane.
(c) Computational grid in the transverse and sagittal plane. 
}
\label{Fig:CFDsketch}  
\end{figure}

A vertiport model is positioned in a uniform flow with a streamwise wind speed $U_{\infty}$. 
The dimensions of the model are shown in Figure~\ref{Fig:CFDsketch} (a). 
The entire domain is represented using a Cartesian coordinate system, with the origin located at the center of the rooftop. 
The $x$-axis aligns with the streamwise direction, defining the wind angle origin ($0^{\circ}$), which increases counterclockwise.
The full-state velocity field data are collected from the rooftop area as a cylindrical region with a radius and height of $5$ m.
Sensors are evenly spaced along the four edges of the vertiport on the rooftop, with a distance of $0.5$ m between each sensor.

The computational domain is discretized into hexahedral elements, with Figure~\ref{Fig:CFDsketch} (b) providing an expanded view of the grid around the vertiport.
For the convenience of setting boundary conditions, the entire domain consists of two main regions: an internal cylinder centered on the vertiport for getting the field data and an external rectangular region, as depicted in Figure~\ref{Fig:CFDsketch} (c). 
The internal cylinder has a radius of $30$ m and a height of $20$ m, with a relatively denser grid distribution to ensure computational accuracy.
The external region primarily serves to define boundary conditions, allowing different wind angles to be simulated by simply rotating the external region. 
The dimensions of the external region are $200$ m, $100$ m, and $20$ m in the $x$, $y$, and $z$ directions, respectively.
The inlet is positioned $80$ m upstream from the center of the vertiport. 
These domain parameters are chosen to balance minimal distortion of the flow structures due to the outer boundary conditions while maintaining computational efficiency.

The fluid flow is simulated by solving the non-dimensionalized incompressible Reynolds-averaged Navier-Stokes equations.
The characteristic scales used for non-dimensionalization are the height of the vertiport $H = 2.8$ m, 
the free stream velocity $U_{\infty}$ 
and twice the dynamic pressure $\rho U_{\infty}^{2}$.
The velocity field is noted $\bm{u}=(u, v, w)$, $p$ is the static pressure, $Re = U_{\infty } H/ \nu$ is the Reynolds number and $\nu$ is the kinematic viscosity.
One RANS simulation is performed for each case, i.e., for each wind condition. 
A commercially available CFD software, ANSYS Fluent 15.0, is employed to discretize the governing equations via the cell-centered Finite Volume Method.
Spatial discretization of the governing equations is conducted using a second-order scheme, while the temporal term is discretized using a first-order implicit scheme.
For the boundary conditions, a uniform streamwise velocity is imposed at the inlet, defined as $\bm{u} = (U_{\infty}, 0, 0)$. 
At the outlet, an outflow condition is applied, which imposes a Neumann condition on the velocity, $\partial_x \bm{u} = (0, 0, 0)$, and a Dirichlet condition on the pressure, $p = 0$. 
A no-slip boundary condition is enforced on the surface of the vertiport.
Additionally, interface boundary conditions are applied to the connecting regions between the inner and outer domains. 
Slip boundary conditions are used on the outer domain, ensuring that any wake-wall interactions are excluded.

To create a CFD database as training data, the wind conditions include nine levels based on the Beaufort scale $b$, with wind angles varying from $0^{\circ}$ to $360^{\circ}$. 
Specifically, the wind speeds are denoted as $b_0 = 0.2$~m/s, $b_1 = 1.5$~m/s, $b_2 = 3.3$~m/s, $b_3 = 5.4$~m/s, $b_4 = 7.9$~m/s, $b_5 = 10.7$~m/s, $b_6 = 13.8$~m/s, $b_7 = 17.1$~m/s, and $b_8 = 20.7$~m/s, with wind angles incremented by $3^{\circ}$, resulting in a total of $1080$ simulation cases.
Four testing datasets are created to evaluate the performance of the model. 
In the first dataset, a single wind speed of $U_\infty = b_4 = 7.9$~m/s is chosen, with all associated wind angles from the CFD database, resulting in a total of $120$ snapshots. 
The second dataset consists of randomly selected snapshots within the CFD database, with multiple wind speeds chosen from $b_0 = 0.2$~m/s to $b_8 = 20.7$~m/s and wind angles from $0^{\circ}$ to $360^{\circ}$, totaling $216$ snapshots. 
Additional simulations have been carried out for the third dataset.
The working conditions are randomly chosen within the database range, i.e., wind speeds are randomly distributed between $0.2$ and $20.7$~m/s and wind angles between $0^{\circ}$ and $360^{\circ}$, totaling $216$ snapshots. 
The fourth dataset contains new simulations with working conditions extending beyond the database range.
Wind speeds are randomly distributed between $0.2$ and $30$~m/s, and wind angles are also randomly distributed between $0^{\circ}$ and $360^{\circ}$,    totaling $216$ snapshots.

\subsection{Verification}
\label{sec2.2}

Before carrying out extensive simulations, grid independence tests are conducted at a wind speed of $U_{\infty} = 13.8$~m/s and a wind angle of $0^{\circ}$ to determine the optimal grid size for the numerical simulation.
Three different grid densities—coarse, medium, and fine—are selected for the inner and outer domains.

\begin{figure}[t]
\includegraphics[width=6.5cm]{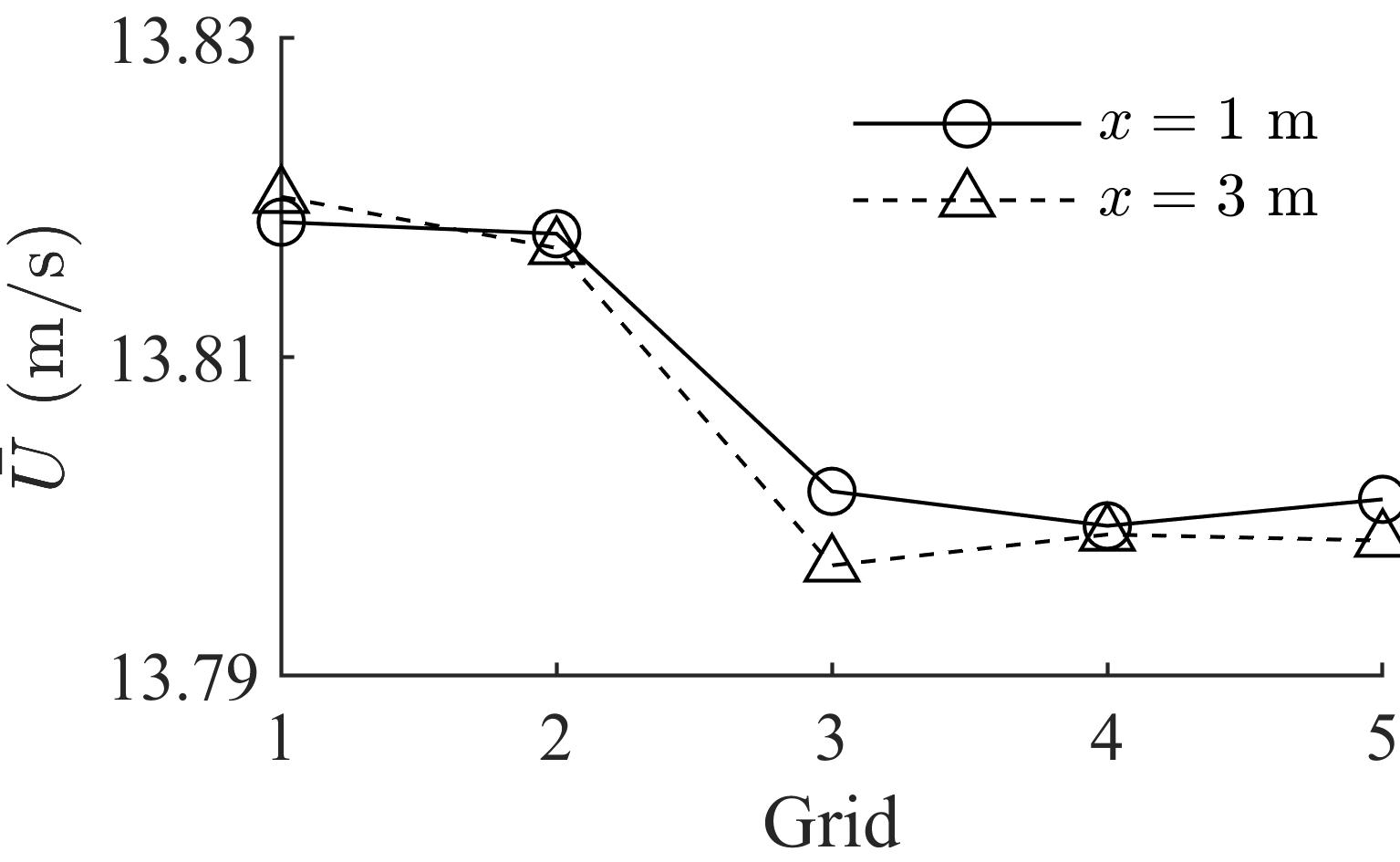}
\caption{Results of the grid independence test conducted using 5 grid sets. The plot shows the mean streamwise velocity $\bar{U}$ across the $y-z$ planes at two downstream locations.}
\label{Fig:Comparison}  
\end{figure}

The validation yields five distinct grid configurations: coarse, coarse outer with medium inner, both medium, medium outer with fine inner, and both fine, as outlined in Table~\ref{Table3}.
\newcommand{\tabincell}[2]{\begin{tabular}{@{}#1@{}}#2\end{tabular}}
\begin{table}[h]
\caption{Grid information for the independence test.}
\label{Table3}
\begingroup
\renewcommand{\arraystretch}{1.1} 
\begin{ruledtabular}
\begin{tabular}{cccc}
\textbf{Grids} & \makecell{\textbf{Inner domain} \\  (\textbf{million})} & \makecell{\textbf{Outer domain} \\  (\textbf{million)}} & \makecell{\textbf{Number of points} \\ (\textbf{million)}} \\[6pt] 
\hline
1  & 362   & 101  & 463  \\  
2  & 543   & 101  & 644  \\  
3  & 543   & 195  & 738  \\  
4  & 543   & 282  & 835  \\  
5  & 734   & 282  & 1016  \\
\end{tabular}
\end{ruledtabular}
\endgroup
\end{table}
To evaluate the grid configurations, we compare the mean streamwise velocity $\bar{U}$ across the $y-z$ planes at two downstream locations ($x = 1$ m, $3$ m), as illustrated in Figure~\ref{Fig:Comparison}.
The first three grid sets show significant variations, indicating that the initial grids lack sufficient resolution. 
However, increasing the grid density beyond the third set results in minimal velocity changes. 
Consequently, the third grid configuration is considered adequate for this study, striking a balance between accuracy and computational efficiency.

\subsection{Flow characteristics}
\label{sec2.3}

\begin{figure}[h]
\includegraphics[width=8.5cm]{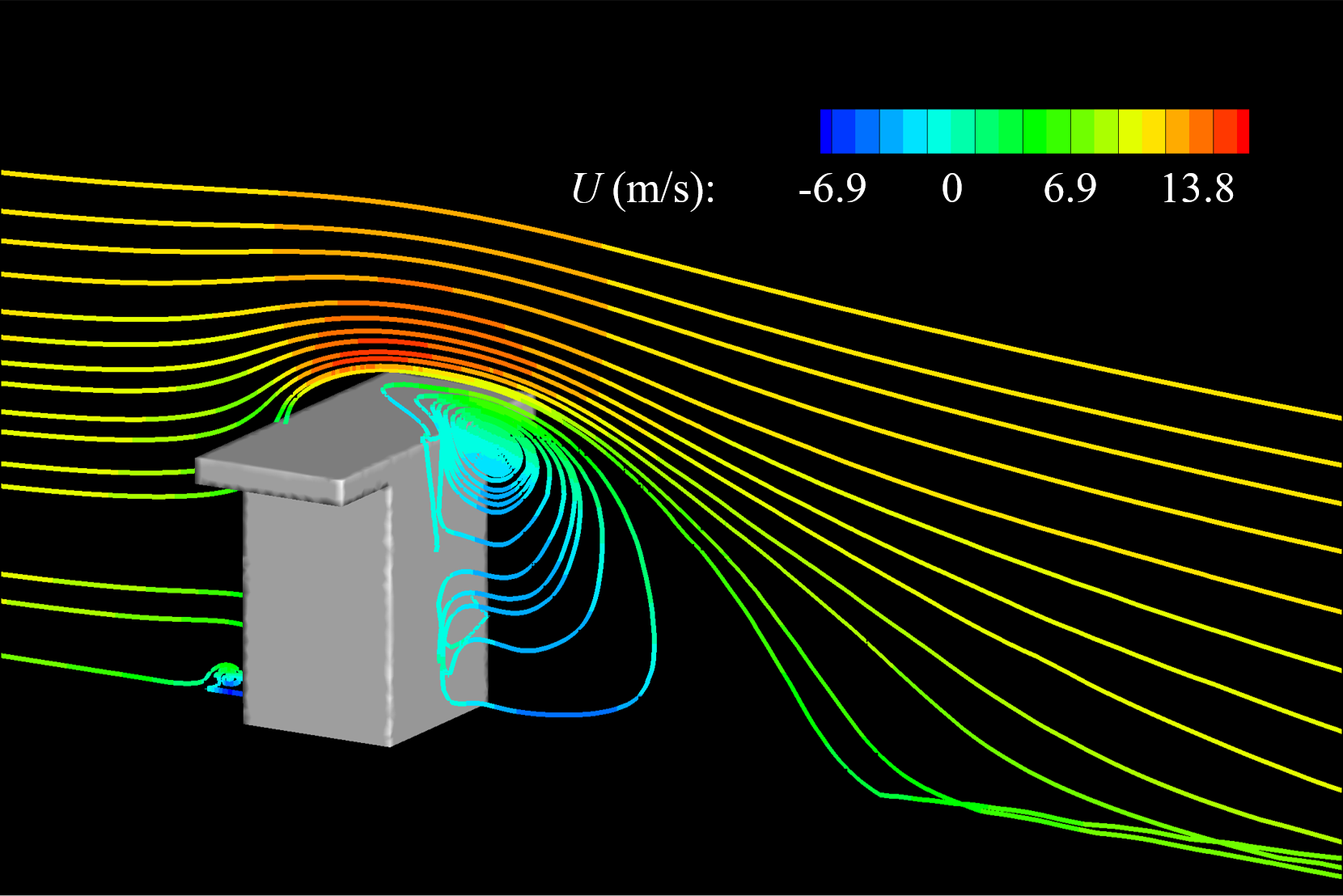}
\caption{Streamlines of the velocity fields in the sagittal plane of the vertiport.}
\label{Fig:flow}  
\end{figure}

Figure~\ref{Fig:flow} illustrates the streamlines of the velocity fields in the sagittal plane of the vertiport, with a wind speed of $U_{\infty} = 13.8$~m/s and wind angle of $\alpha = 0^{\circ}$.
The streamlines are color-coded by the streamwise velocity, exhibiting the flow characteristics around the vertiport.
Near the rooftop, the streamwise velocity exhibits significant variation: flow reversal around the vertiport is observed, with a wide recirculation zone. The far-field above experiences notable acceleration, resulting in the formation of a shear layer away from the vertiport. 
These flow patterns are crucial for drone operations, particularly during take-off and landing. 
The reversed flow can displace drones and the shear layer may compromise their stability. 
Therefore, an accurate estimation of the velocity field is essential for developing effective control algorithms, with a direct impact on drone safety.

\section{Machine-learned flow estimation with sparse data}
\label{sec3}

\begin{figure}
\includegraphics[width=8.5cm]{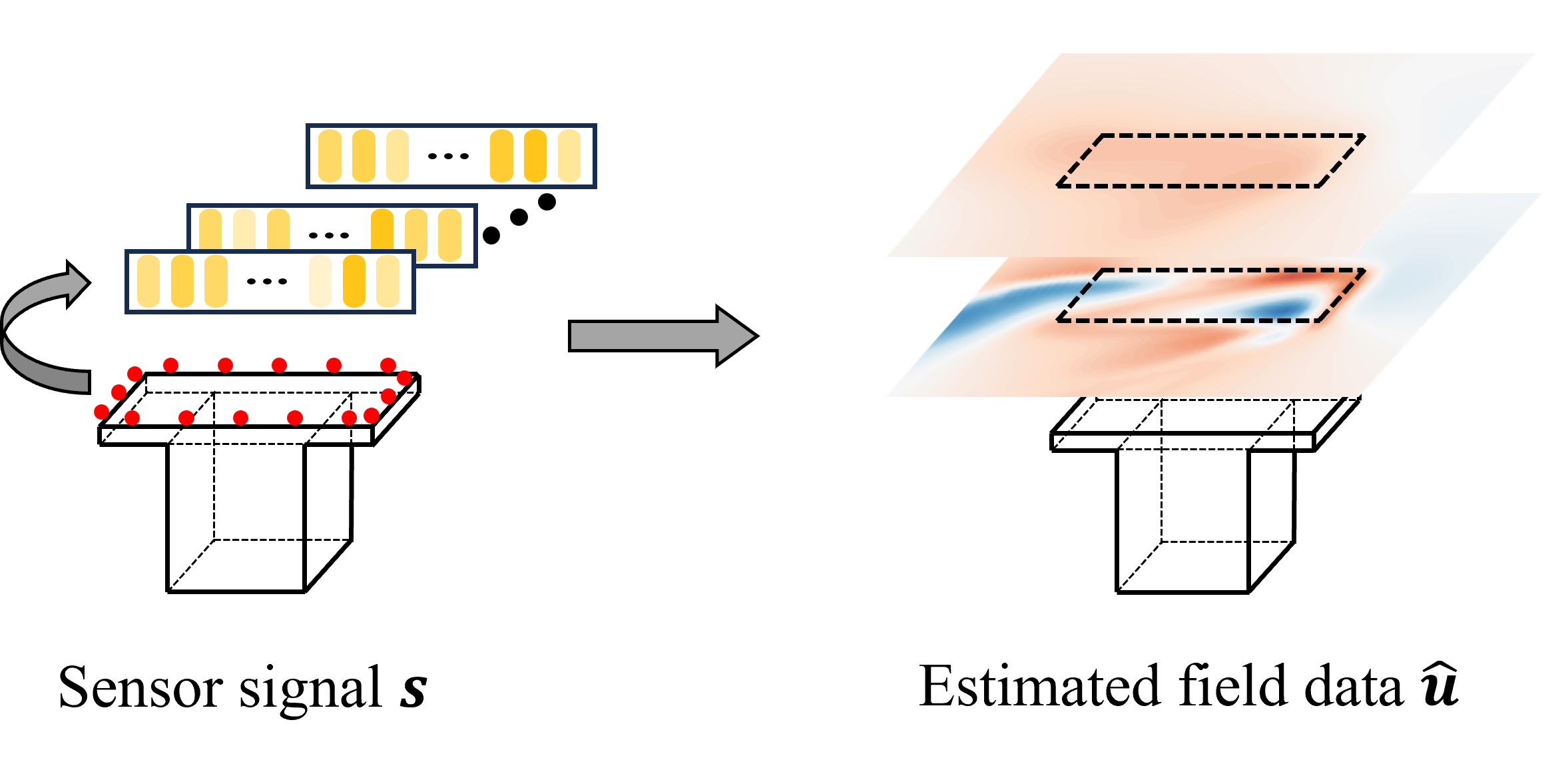}
\includegraphics[width=8.5cm]{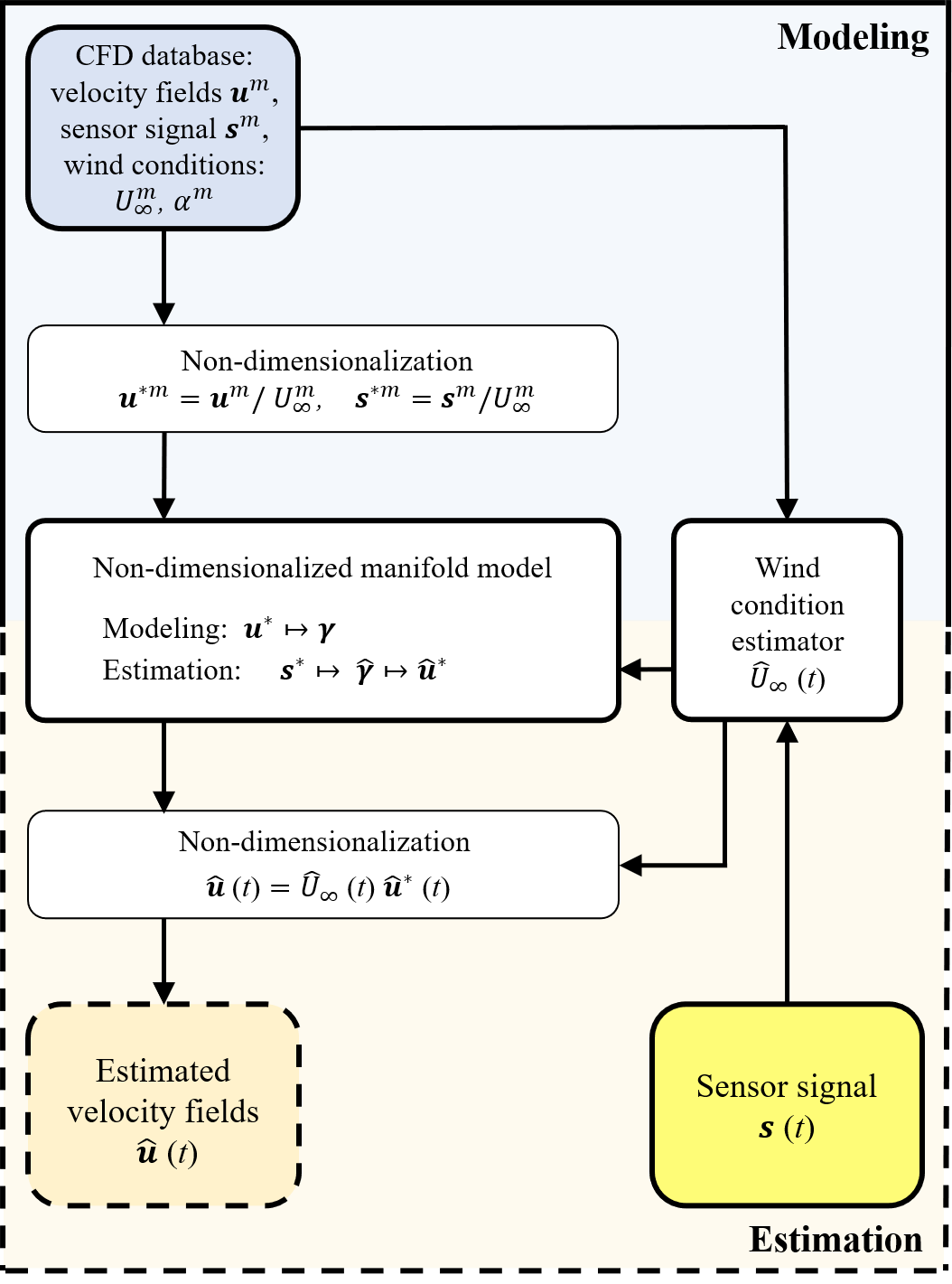}
\caption{Methodology framework.
Modeling: The dataset is first non-dimensionalized based on the oncoming wind speed. 
The non-dimensionalized velocity fields serve as training inputs for the manifold model, using the ISOMAP-$k$NN manifold learner.
Estimation:  The wind condition estimator first predicts the oncoming wind speed for a given testing sensor signal with index $t$. 
The sensor signals are then non-dimensionalized based on this estimated speed and input into the manifold learner. 
After mapping these signals to latent variables, the manifold model reconstructs the complete velocity fields.}
\label{Fig:Methodology}  
\end{figure}
\begin{figure*}
\includegraphics[width=12cm]{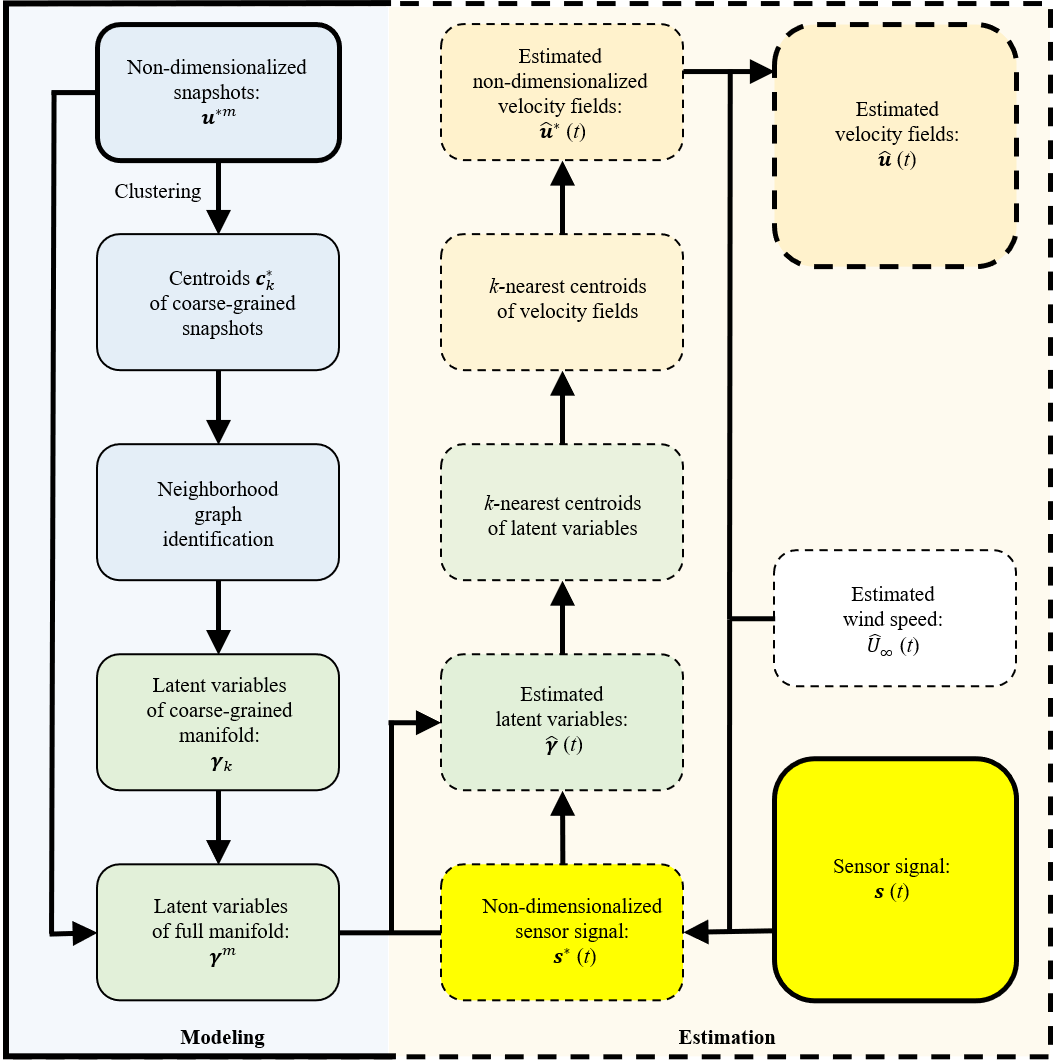}
\caption{Non-dimensionalized ISOMAP-$k$NN:  
Clustering is introduced to reduce computational load. 
In the modeling part, the coarse-grained manifold is derived from the cluster centroids, while the snapshot manifold is obtained through a $k$NN-based projection. 
For estimation, the latent variables from the non-dimensionalized sensor signals are employed to reconstruct the velocity field via a first-order Taylor expansion based on the velocity fields of the centroids.}
\label{Fig:cISOMAP}  
\end{figure*}
\begin{figure}
\includegraphics[width=6.5cm]{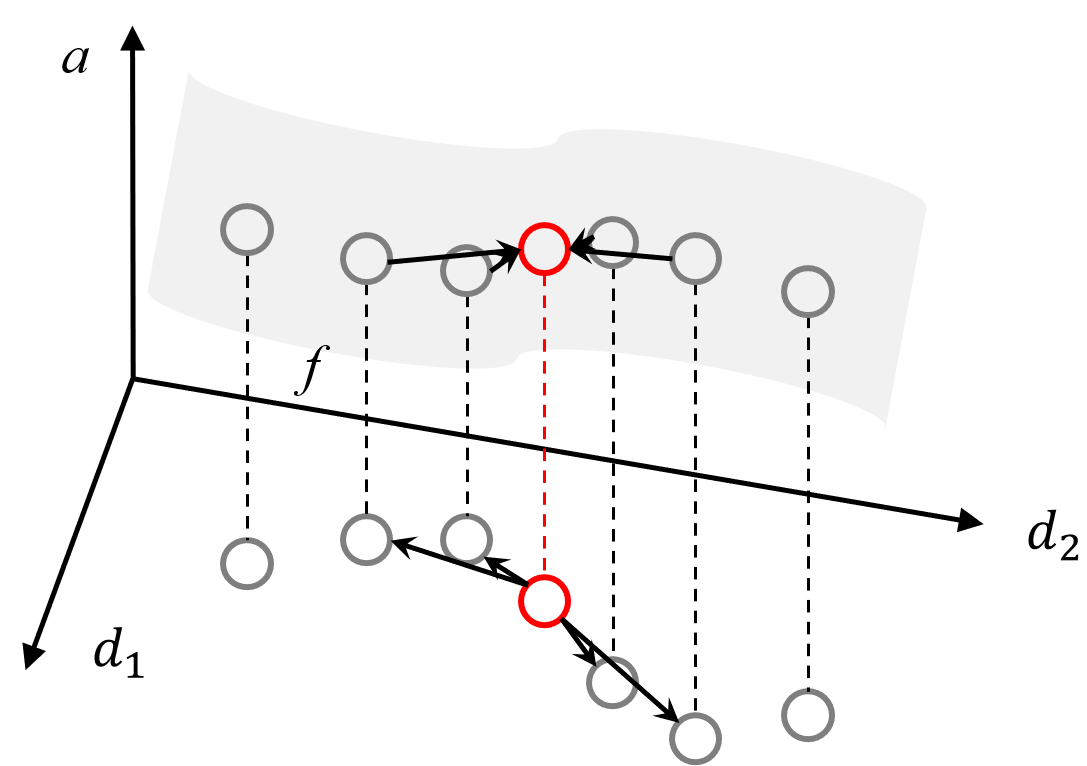}
\caption{A schematic of $k$NN. $d_1$ and $d_2$ are the original feature space, $f$ is a mapping function, and $\bm{a}$ denotes the
output feature.
Example of a point (red) and its nearest neighbors (connected with arrows).}
\label{Fig:KNN}  
\end{figure}
In this section, we present the framework for machine-learned flow estimation with sparse data,  as illustrated in Figure~\ref{Fig:Methodology}. 
The framework comprises two main components: modeling and estimation. 
In the modeling phase, the dataset, which includes the velocity field and sensor signals, is non-dimensionalized with the oncoming wind speed $U_{\infty}$.
The non-dimensionalized velocity fields serve as training inputs for the manifold model. 
The objective is to establish a mapping between the high-dimensional field data and the low-dimensional latent variables from the manifold employing the ISOMAP-$k$NN~\citep{tenenbaum2000global} manifold learner.
ISOMAP-$k$NN is well-suited for providing accurate low-dimensional representations across a wide range of operating conditions~\citep{farzamnik2023snapshots}.
In the estimation phase, when there is a testing sensor signal, the process begins with a wind condition estimator that provides the oncoming wind speed.
The sensor signals are then non-dimensionalized based on this estimated speed and input into the manifold learner. 
After mapping these signals to latent variables, the decoder is conducted to reconstruct the complete velocity field.
This approach facilitates accurate extrapolation under rare wind conditions by leveraging non-dimensionality within the manifold model.

Section \S \ref{sec3.1} describes the non-dimensionalization of the dataset, while \S \ref{sec3.2} details the clustering approach. 
The manifold learner is discussed in \S \ref{sec3.3}, and finally, sensor-based flow estimation is addressed in \S \ref{sec3.4}.

\subsection{Non-dimensionalization of the data}
\label{sec3.1}

The velocity fields and sensor signals are non-dimensionalized using the characteristic wind speed $U_\infty$. 

For the field data $\bm{u}$, the non-dimensionalization is defined as:
\begin{equation}
 \bm{u}^*  = \bm{u}/U_\infty.
\end{equation}
Similarly, for the velocity sensor data $\bm{s}$, the non-dimensionalization is:
\begin{equation}
\bm{s}^* = \bm{s}/U_\infty.
\label{eq:Vscaling}
\end{equation}
These non-dimensionalized data, $u^*$ and $s^*$, are used as inputs for the manifold model, facilitating a more general analysis across different working conditions.

\subsection{Clustering}
\label{sec3.2}

Clustering is used to coarse-grain the existing dataset into several groups based on their similarities.
Each cluster is represented by its centroid, allowing operations performed on the data snapshots to be applied to the centroids instead. 
In this study, the unsupervised $k$-means++ algorithm is used for clustering. 
This algorithm aims to maximize similarity within clusters and minimize similarity among clusters \citep{2014_JFM_kaiser_cluster}.
For an ensemble of $M$ statistically representative snapshots $\bm{u}^{m}$, with $m = 1, \ldots , M$, the snapshots are coarse-grained into $K_c$ clusters, noted $\mathcal{C}_k$, with $k = 1, \ldots , K_c$.
The center of mass of the clusters is defined as its centroid noted $\bm{c}_{k}$.

We first define a Hilbert space,
for two square-integrable functions $\bm{u}$ and $\bm{v}$,  the inner product in domain $\Omega$ is:
\begin{equation}
(\bm{u}, \bm{v}) _\Omega = \int_{\Omega} \bm{u} \cdot \bm{v} \ \mathrm{d} x .
\end{equation}
The distance is measured using the norm $\left\| \cdot \right\|_{\Omega}$, defined as:
\begin{equation}
\|\bm{u}\|_{\Omega} = \sqrt{(\bm{u}, \bm{u})_\Omega}.
\end{equation}

The unsupervised $k$-means++ algorithm iteratively updates the optimal location of the cluster centroids according to the data distribution in the state space:
\begin{enumerate}
    \item Initialization:
    The $K_c$ centroids are selected from the snapshots. 
    Compared to the $k$-means algorithm, the $k$-means++ algorithm optimizes the initial centroids to avoid local optimality and speed up convergence.
    
    \item Assignment:  The corresponding cluster-affiliation function maps the snapshot to the index of the closest centroid:
    \begin{equation}
    k(m)=\arg \min _{i}\left\|\bm{u}^m-\bm{c}_{i}\right\|_{\Omega}
    \end{equation}
    Alternatively, the characteristic function $\chi_{k}^{m}$:
    \begin{equation}
    \chi_{k}^{m}:=\left\{\begin{array}{ll}
    1, & \text { if } \bm{u}^{m} \in \mathcal{C}_{k} \\
    0, & \text { otherwise }
    \end{array}\right.
    \label{eq4}
    \end{equation}
    describes whether the $m$-th snapshot is affiliated with the $k$-th centroid. 
    
    \item Update: 
    The original cluster centroids $\bm{c}_{k}$ are updated by averaging all the snapshots within cluster $\mathcal{C}_k$:
    \begin{equation}
    \bm{c}_{k}=\frac{1}{n_{k}} \sum_{m=1}^{M}  \chi_{k}^{m} \bm{u}^{m}, 
    \end{equation}
    where $n_{k} =  \sum_{m=1}^{M}{\chi_{k}^{m}}$ is the number of snapshots in cluster $\mathcal{C}_k$. 

    \item Iteration:
    The assignment and update steps are repeated until convergence is reached.
    Convergence means that the centroids do not vary or stabilize below a certain threshold. 
    The algorithm minimizes the intra-cluster variance and maximizes the inter-cluster variance. 
    The intra-cluster variance $J$ of this cluster partition is computed as:
    \begin{equation}
    J\left(\bm{c}_{1}, \ldots, \bm{c}_{K_c}\right)= \frac{1}{M}  \sum_{m=1}^{M} \left\|\bm{u} ^{m}-\bm{c}_{k(m)}\right\|_{\Omega}^2.
    \end{equation}
\end{enumerate}
Based on \citet{2021_JFM_li_cluster}, we set the parameters for the $k$-means++ algorithm to $1000$ iterations and $100$ repetitions to ensure optimal partitioning.
The averaged centroids $\bm{c}_{k}$ and the cluster affiliation $k({m})$ of the snapshots are taken as the final results.

The computational cost of clustering algorithms is also taken into account for large, high-dimensional datasets.
Optional pre-processing can be performed using a \emph{lossless} POD method to perform data compression \citep{2014_JFM_kaiser_cluster,2021_JFM_li_cluster,2021_SCIA_fernex_cluster,2022_JFM_deng_cluster}.
The snapshot $\bm{u}^{m}$ can be expressed by the POD expansions as:
\begin{equation}
\bm{u}^{m} - \bm{u}_{0} \approx \sum_{i=1}^{N} a_{i}^{m}\bm{u}_{i},
\end{equation}
where $\bm{u}_{0}$ is the constant mean of the dataset, 
$\bm{u}_{i}$ are the spatial POD modes, with the corresponding mode coefficients $\bm{a}^{m} = \left[a_{1}^{m}, \ldots, a_{N}^{m}\right]^{\top}$ for different snapshots. 
A complete basis for the decomposition can be expected when $N = M$.
Thus, the distance can be computed with $\bm{a}^{m}$ instead of the full-dimensional snapshots, i.e.,
\begin{equation}
\left\|\bm{u}^{m}-\bm{u}^{n}\right\|_{\Omega} =\left\|\bm{a}^{m}-\bm{a}^{n}\right\|_2.
\end{equation}
where $\Vert \cdot \Vert_2$ denotes the Euclidean norm.

The computational load of clustering is significantly reduced by applying to the $N$-dimensional state vector $\bm{a}^{m}$ instead of performing an expensive computation on the full domain with millions of grid nodes. 
We emphasize that the POD is only an optional preprocessing to speed up the computation and is not necessary for the clustering algorithm.

\subsection{Non-dimensionalized manifold learner}
\label{sec3.3}
We develop a non-dimensionalized manifold learner based on the approach of ~\citet{farzamnik2023snapshots}, as shown in figure~\ref{Fig:cISOMAP}. 
The proposed manifold learner consists of two main parts. 
First, in the modeling part, the high-dimensional data is embedded into a low-dimensional space using isometric feature mapping. 
This data-driven encoding process uncovers a hidden manifold, allowing us to connect the new coordinates to physical features of the dataset, such as wind angle and wind speed around the vertiport.
Subsequently, a decoding process is included in the estimation part, enabling the latent variables to return to the high-dimensional space and reconstruct the original velocity fields.

\subsubsection{Encoder}
The $k$-nearest neighbor regression algorithm~\citep{stone1977consistent,fix1985discriminatory} is the foundation for constructing the manifold model.
The $k$NN algorithm is a non-parametric approach to approximate a mapping function $f$ by averaging the outputs of the $k$-nearest observations to the input data, as shown in figure~\ref{Fig:KNN}. 
Choosing a smaller $k$ can make the algorithm sensitive to outliers and noise, while a larger $k$ generally increases the computational cost.
For a velocity field $\bm{u}^m$, we define $\mathcal{N}(\bm{u}^m; K, \bm{u}^1, \ldots, \bm{u}^M) $ as the $K$ nearest neighbors of this velocity field in the dataset $\bm{u}^1, \ldots, \bm{u}^M$.

We enhance this manifold learner by incorporating a clustering procedure. 
The resulting cluster-based ISOMAP (cISOMAP) uses the centroids of the clusters instead of the original snapshots. 
This coarse-grained approach reduces computational load while maintaining model accuracy. 
The additional representation error introduced by the clustering procedure will be discussed in the next section.

ISOMAP is a nonlinear dimensionality reduction technique that creates a low-dimensional embedding. 
This embedding is designed to preserve the geodesic distances between data points in the original high-dimensional space.
For a set of $M$ non-dimensionalized snapshots $\bm{u}^{*m}$, the $k$-means++ algorithm groups them into $K_c$ clusters, each represented by a centroid  $\bm{c}^*_k$, with $k = 1, \ldots, K_c$.
These centroids serve as the foundation for constructing the neighborhood graph through the following steps.
First, the norm $\left\| \bm{c}^*_{i}-\bm{c}^*_{j} \right\|_{\Omega}$ between each pair of centroids $\bm{c}^*_{i}$ and $\bm{c}^*_{j}$, is computed for all centroids. 
Next, a neighboring graph $\bm{G}$ is constructed, connecting each pair of nodes (centroids) $\bm{c}^*_{i}$ and $\bm{c}^*_{j}$ by an edge with a weight of $\left\| \bm{c}^*_{i}-\bm{c}^*_{j} \right\|_{\Omega}$,
if $\bm{c}^*_j \in \mathcal{N}(\bm{c}^*_{i}; K_e, \bm{c}^*_1, \ldots, \bm{c}^*_{K_c}) $. 
Here, $\mathcal{N}(\bm{c}^*_{i}; K_e, \bm{c}^*_1, \ldots, \bm{c}^*_{K_c}) $ denotes the $K_e$ nearest neighbors of $\bm{c}^*_{i}$ from all the centroids $\bm{c}^*_1, \ldots, \bm{c}^*_{K_c}$.
Finally, this norm $\left\| \bm{c}^*_{i}-\bm{c}^*_{j} \right\|_{\Omega}$ between all pairs of snapshots in $\bm{G}$ are replaced by the geodesic distances 
calculated using Floyd’s algorithm~\citep{floyd1962algorithm,farzamnik2023snapshots}. 
These geodesic distances are stored in the matrix $\bm{D}_G$. 

The geodesic distances are further used as input for classical multidimensional scaling (MDS)~\citep{torgerson1952multidimensional} to construct the low-dimensional embedding.
This process ensures that the Euclidean pairwise distances between the latent variables in the reduced space approximate the geodesic distances.
The MDS algorithm seeks to find a matrix $\bm{\Gamma}$ that minimizes the following cost function:
\begin{equation}
    \left \| \bm{\Gamma} \bm{\Gamma} ^\top -  \bm{B} \right \| _F^2,
    \label{eq:MDS}
\end{equation}
where $\bm{B}=-\frac{1}{2} \bm{H}^{\top}\left(\bm{D}_G \odot \bm{D}_G\right) \bm{H}$,
with $\bm{H} = \bm{I}_N  -  (1 / N) \mathds{1}_{N}$ being the centering matrix, $\bm{I}_N$ the identity matrix of dimension $N$, $\mathds{1}_{N}$ the all-ones matrix of dimension $N$, $\odot$ the Hadamard (element-wise) product and $\left \|  \cdot \right \|_F$ represents the Frobenius norm\citep{farzamnik2023snapshots}. 
Given the eigendecomposition of $\bm{B} = \bm{V} \bm{\Lambda} \bm{V}^\top$, with $\bm{V}$ being the eigenvector matrix and $\bm{\Lambda}$ the eigenvalue matrix, we have $\bm{\Gamma}=\bm{\Lambda}^{(1/2)}\bm{V}$.
For a low-dimensional embedding of dimension $p$, only the first $p$ columns of $\bm{\Gamma}$, $\bm{\bar{\gamma}}_1, \dots, \bm{\bar{\gamma}}_p$, are considered.

The coarse-grained manifold is obtained based on $\bm{c}^*_k$, with the latent variables defined as $\bm{\bar{\gamma}}_k$.
We use the $k$NN regression with all the velocity field data to obtain the full manifold.
The corresponding latent variables for the full manifold can be expressed as:
\begin{equation}
\bm{\gamma}^m = \sum_{k = 1}^{K_c} \bm{w}_k^{*} (m) \bm{\bar{\gamma}}_k,
\end{equation}
where $\bm{w}_k^{*} (m)$ is the corresponding weight of the latent variable $\bm{\bar{\gamma}}_k$ in the coarse-grained manifold.
We define a characteristic function $\chi^*_m (k)$ by:
\begin{eqnarray}
\chi^*_m (k):=
\left\{\begin{array}{ll}
1, & \text { if } \bm{c}^*_{k} \in \mathcal{N}(\bm{u}^{*m}; K_p, \bm{c}^*_1, \ldots, \bm{c}^*_{K_c})  \\
0, & \text { otherwise. }
\end{array}\right.
\\
\sum_{k = 1}^{K_c} \chi^*_m (k) = K_p,
\end{eqnarray}
where $\mathcal{N}(\bm{u}^{*m}; K_p, \bm{c}^*_1, \ldots, \bm{c}^*_{K_c})$ is the set of $K_p$ nearest neighbours to $\bm{u}^{*m}$ in $\bm{c}^*_1, \ldots, \bm{c}^*_{K_c}$.
$\bm{w}_k^{*} (m)$ is then defined as:
\begin{equation}
\bm{w}_k^{*} (m) = \frac{\chi^*_m (k) /\left \| \bm{u}^{*m} - \bm{c}^*_k  \right \|_{\Omega} }
{\sum_{i = 1}^{K_c}  \chi^*_m (i) /\left \| \bm{u}^{*m} - \bm{c}^*_i  \right \|_{\Omega}}.
\end{equation}
%

To evaluate the effectiveness of ISOMAP, \citet{tenenbaum2000global} introduced the concept of residual variance, defined as:
\begin{equation}
R = 1- r^2(\mathrm{vec} (\mathbf{D}_G),\mathrm{vec} (\mathbf{D}_{\bm{\gamma}})).
\end{equation}
Here,
$\mathbf{D}_{\bm{\gamma}}$
represents the matrix of norm between points in the low-dimensional embedding, $r^2$ is the squared correlation coefficient, and ‘vec’ denotes the vectorization operator. 
The residual variance ranges from $0$ to $1$, indicating the proportion of information not captured by the low-dimensional embedding of the original data.

\subsubsection{Decoder}
We also employ a data-driven approach for the decoder, which links the latent variables to the high-dimensional velocity fields.
Each centroid $\bm{c}^*_{k}$ has a corresponding low-dimensional counterpart $\bm{\bar{\gamma}}_k$, where $k = 1, \dots, K_c$. 
Given a low-dimensional representation $\bm{\gamma} (t)$ as input, the $k$NN algorithm is used to find the nearest latent variables $\bm{\gamma}_{(i)}$ within the centroid manifold, where $(i)$ stands for the $i$-th nearest neighbors, followed by linear interpolation to transfer the low-dimensional representation into the snapshot data.

Let $f$ be the unknown mapping that converts the latent variables back into the high-dimensional space. 
To reconstruct the flow field for any $\bm{\gamma} (t)$, we first identify its $K_d$-nearest neighbors $\bm{\bar{\gamma}}_{(1)}$, $\dots$ , $\bm{\bar{\gamma}}_{(K_d)}$ and their corresponding high-dimensional counterparts $\bm{c}^*_{(1)}$, $\dots$, $\bm{c}^*_{(K_d)}$. 
Typically, setting $K_d = K_p$ yields reasonable reconstruction accuracy.
The reconstruction (decoding) of $\bm{\gamma} (t)$ is then approximated by a first-order Taylor expansion starting from the nearest neighbor, $\bm{c}^*_{(1)}$, mapping back to the original space:
\begin{equation}
    \bm{u}^* (t) = \bm{c}^*_{(1)}+(\bm{\gamma} (t)-\bm{\bar{\gamma}}_{(1)})\nabla f(\bm{\bar{\gamma}}_{(1)})^\top.
\end{equation}
where the gradient tensor at $\bm{\bar{\gamma}}_{(1)}$ is given by:
\begin{equation}
    \nabla f(\bm{\bar{\gamma}}_{(1)}) = ({\partial f}/{\partial \bm{\gamma}_{1} (\bm{\bar{\gamma}}_{(1)})} ,\dots, {\partial f}/{\partial \bm{\gamma}_{p} (\bm{\bar{\gamma}}_{(1)})}).
\end{equation}
This tensor is estimated by assuming an orthogonal projection of the $K_d - 1$ directions from the low-dimensional embedding to the original feature space, as follows:
\begin{equation}
    \left[\begin{array} {c} \bm{c}^*_{(2)}-\bm{c}^*_{(1)} \\ \cdots \\ \bm{c}^*_{(K_d)}-\bm{c}^*_{(1)} \end{array}\right] \simeq\left[\begin{array}{c} \bm{\bar{\gamma}}_{(2)}-\bm{\bar{\gamma}}_{(1)} \\ \cdots \\ \bm{\bar{\gamma}}_{(K_d)}-\bm{\bar{\gamma}}_{(1)} \end{array}\right] \nabla f\left(\bm{\bar{\gamma}}_{(1)}\right)^{\top}
    \label{eq:Taylor}
\end{equation}
which leads to the approximation
$\nabla f(\bm{\bar{\gamma}}_{(1)})^\top =
(\triangle \mathbf{G}^\top  \triangle  \mathbf{G} )^{-1}\triangle \mathbf{G} ^\top\triangle \mathbf{C}$ using least squares minimization, where $\mathbf{C}$ and $\mathbf{G} $
represent the left-hand side and the first term on the right-hand side of equation \eqref{eq:Taylor}, respectively. 

The ISOMAP algorithm described here allows for various choices of norms other than the Euclidean norm, different methods for identifying neighbors to construct the neighboring graph $\bm{G}$, other shortest path algorithms, or even a non-classical approach to MDS~\citep{farzamnik2023snapshots}.

\subsection{Flow estimation from sensor signal}
\label{sec3.4}
The flow estimation begins with a wind condition estimator that determines the wind speed $\hat{U}_\infty$ from a testing sensor signal $\bm{s}(t)$. 
When the testing data is far beyond the existing database, extrapolation becomes essential. 
Here, we implement a scalable estimator that effectively handles the extrapolation process.

First, a scaling factor $\lambda^l (t)$ is determined by minimizing:
\begin{eqnarray}
\lambda^m (t) = \arg \underset{\mu}{\min} \left\| \mu \bm{s}^{m} - \bm{s}(t) \right\|_{2},\\
l (t) = \arg \underset{m}{\min} \left\| \lambda^m (t) \bm{s}^{m} - \bm{s}(t) \right\|_{2}. 
\label{eq:Vsestimation}
\end{eqnarray}
In the first equation, for each sensor $\bm{s}^m$ in the database, the corresponding scaling factor $\lambda^m$ is determined. 
Then, $\lambda^l (t)$ is found by selecting the $l (t)$-th $\lambda$ with the minimum error from evaluating all $\lambda^m$ in the second equation.

The estimated wind speed is obtained by:
\begin{equation}
\hat{U}_\infty (t) = \lambda^l (t) U_\infty^{l},
\end{equation}
where $U_\infty^{l}$ is the corresponding wind speed of the referred sensor signal $\bm{s}^{l}$ from the database.

Consequently, the sensor signal is normalized with $\hat{U}_\infty(t)$:
\begin{equation}
\bm{s}^*(t) = \bm{s}(t)/\hat{U}_\infty(t).
\label{eq:Vstcaling}
\end{equation}
The estimated latent variables can be obtained by:
\begin{equation}
\hat{\bm{\gamma} }(t) = \sum_{m = 1}^{M} \bm{w}^*_m(t) \bm{\gamma}^m,
\end{equation}
where $\bm{w}^*_m(t)$ is the weight of the latent variable $\bm{\gamma}^m$.
The characteristic function is defined as: 
\begin{eqnarray}
\chi^*_{t}(m):=
\left\{\begin{array}{ll}
1, & \text { if } \bm{s}^{*m} \in \mathcal{N}(\bm{s}^* (t); K_w, \bm{s}^{*1}, \ldots, \bm{s}^{*M}) \\
0, & \text { otherwise. }
\end{array}\right.
\\
\sum_{m = 1}^{M} \chi^*_{t}(m) = K_w,
\end{eqnarray}
Where $\mathcal{N}(\bm{s}^* (t); K_w, \bm{s}^{*1}, \ldots, \bm{s}^{*M})$ is the set of $K_w$ nearest neighbors to $\bm{s}^* (t)$ in $\bm{s}^{*1}, \ldots, \bm{s}^{*M}$.
$\bm{w}^*_m(t)$ is subsequently defined as:
\begin{equation}
\bm{w}^*_m(t) = \frac{\chi^*_{t}(m) /\left \| \bm{s}^*(t) - \bm{s}^{*m}  \right \|_{2} }{\sum_{i = 1}^{M}  \chi^*_{t}(i) /\left \| \bm{s}^*(t) - \bm{s}^{*i}  \right \|_{2}}.
\end{equation}

The latent variable $\hat{\bm{\gamma}}(t)$ is delivered into the manifold decoder, yielding the non-dimensionalized estimated velocity field $\hat{u}^*(t)$.
Compared to the traditional $k$NN approach, which directly predicts the flow field from sensor signals, the embedding of manifold models in the ISOMAP-$k$NN approach introduces latent variables as intermediaries, offering larger physical interpretability.
The estimated velocity field \( \hat{u}(t) \) is determined by rescaling:
\begin{equation}
    \hat{u}(t) = \hat{U}_\infty (t) \hat{u}^*(t).
\end{equation}
%

\section{Results}
\label{sec4}
\begin{figure}
\includegraphics[width=6.5cm]{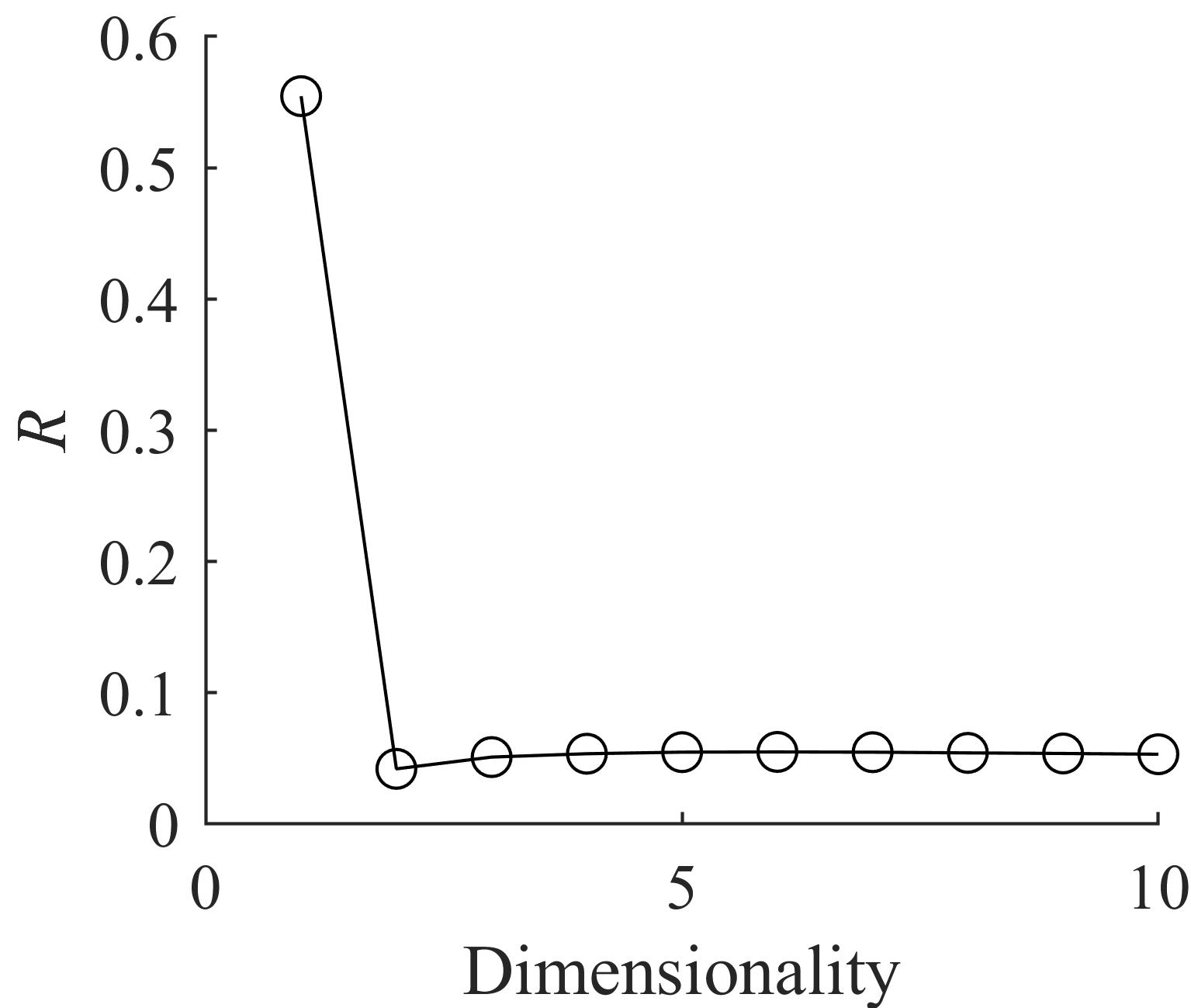}
\caption{
Residual variance $R$ of the low-dimensional embedding as a function of the dimensionality for the coarse-grained centroid manifold.
}
\label{Fig:choiceK}  
\end{figure}
\begin{figure*}
\includegraphics[width=15cm]{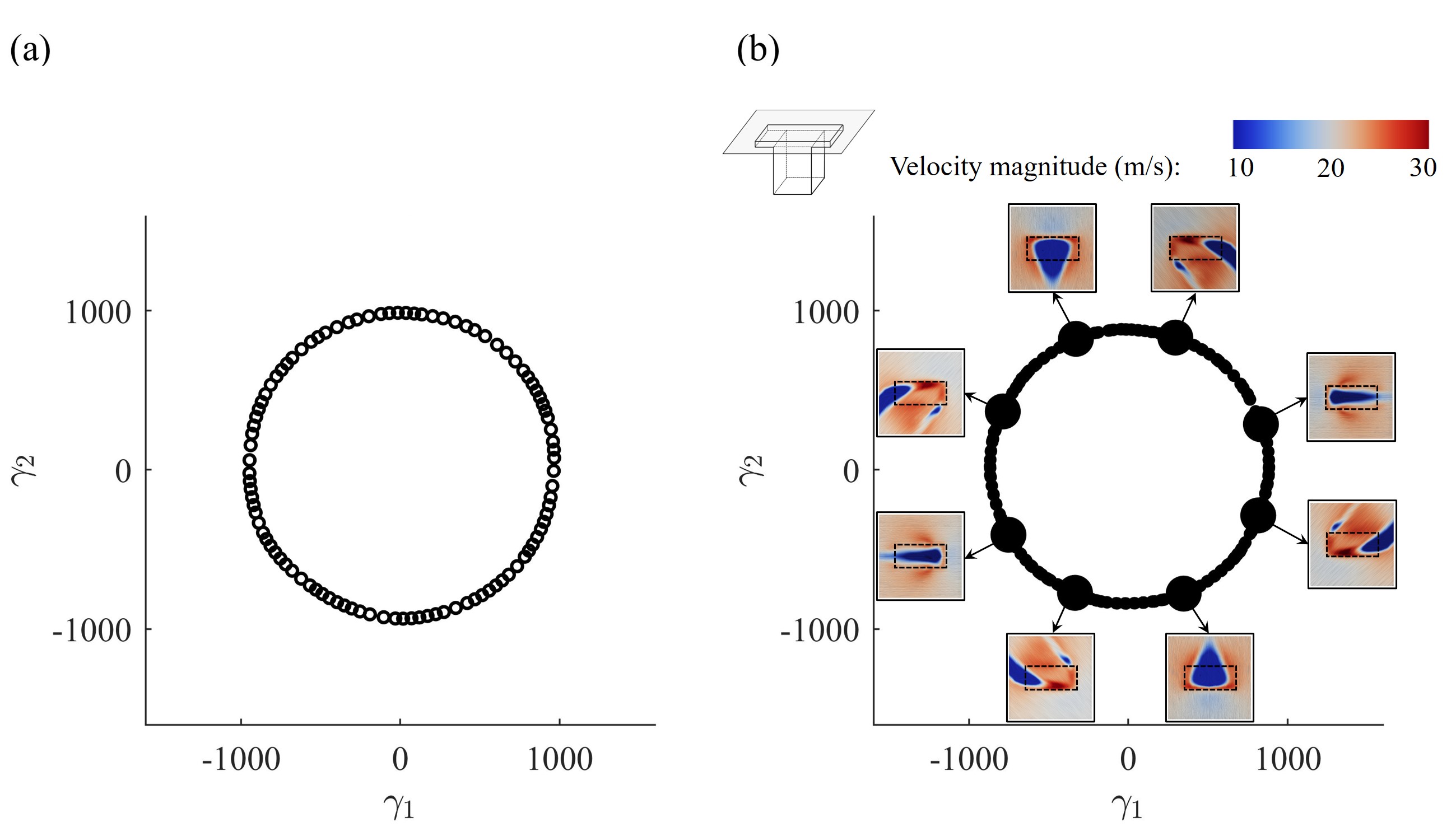}
\caption{
(a) The centroid-based manifold and (b) the snapshot-based manifold, with the representative flow state of $U_\infty = 20.7$~m/s at $z = 0.2$~m visualized around the manifold.
$\gamma_1$ and $\gamma_2$ represent the latent variable coordinates.
There is a slight shift in the current feature coordinates with respect to the wind angle, causing the distribution to be asymmetric. 
This shift is determined by the MDS process in the encoder of the manifold leaner, which focuses on the manifold shape and may generate the latent variables on rotated feature coordinates if the manifold is axisymmetric.
}
\label{Fig:manifolds}  
\end{figure*}

\begin{figure*}
\includegraphics[width=7.5cm]{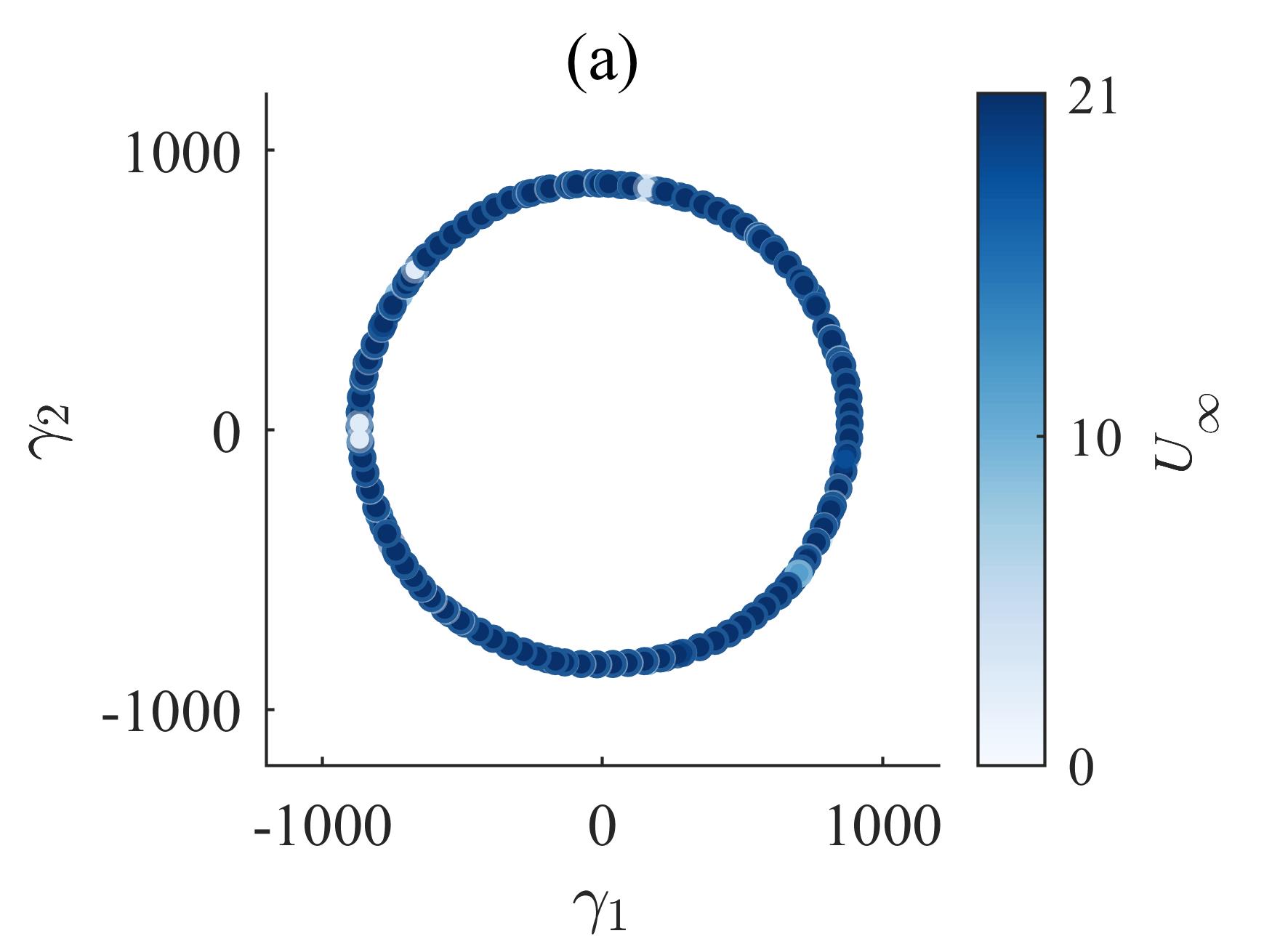}
\includegraphics[width=7.5cm]{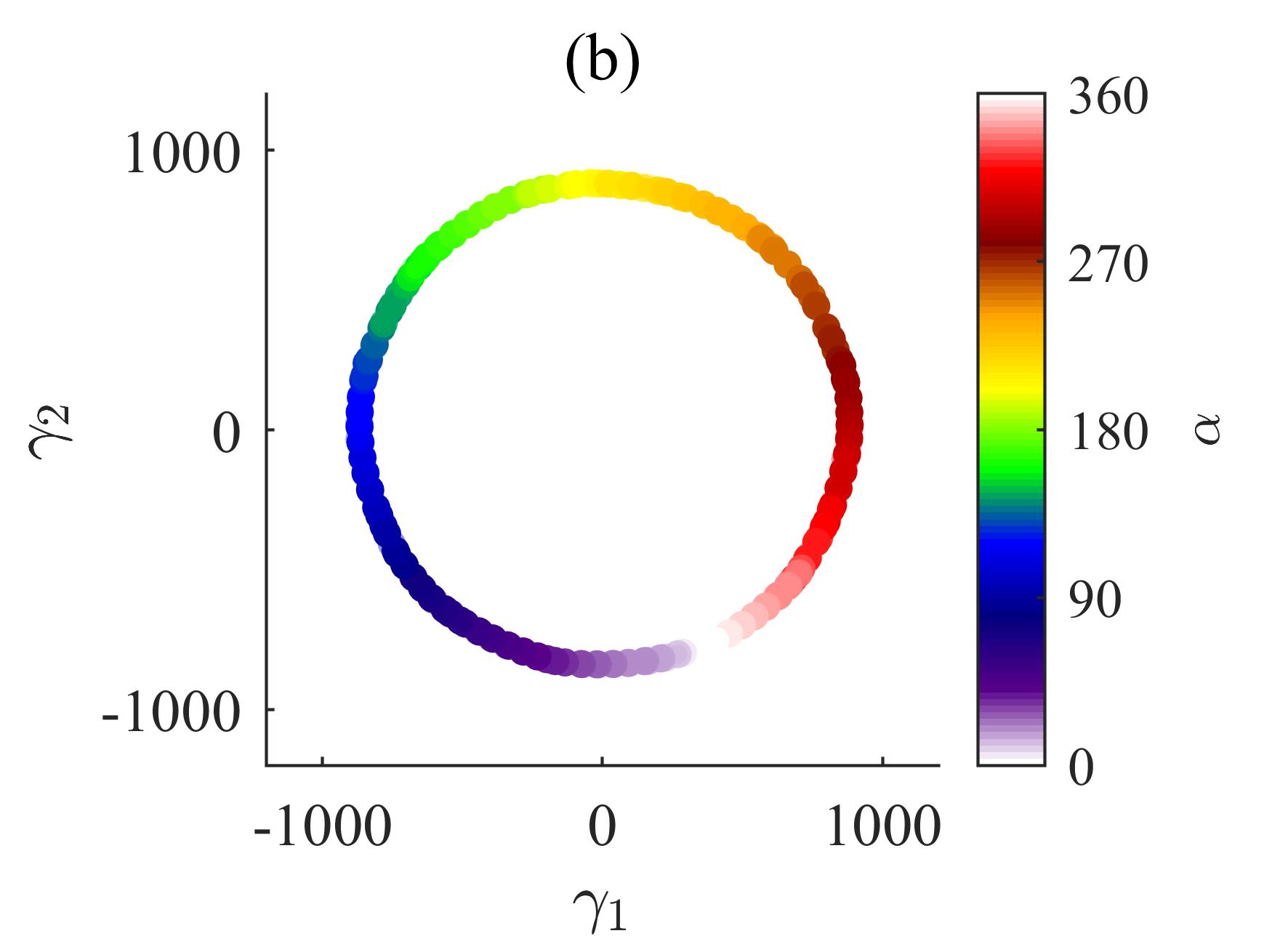}
\caption{The latent variable $\bm{\gamma}$ and the wind conditions on the snapshot manifold. (a) Wind speed. (b) Wind angle.}
\label{Fig:manifoldsWindConditions}  
\end{figure*}
\begin{figure}
\includegraphics[width=6.5cm]{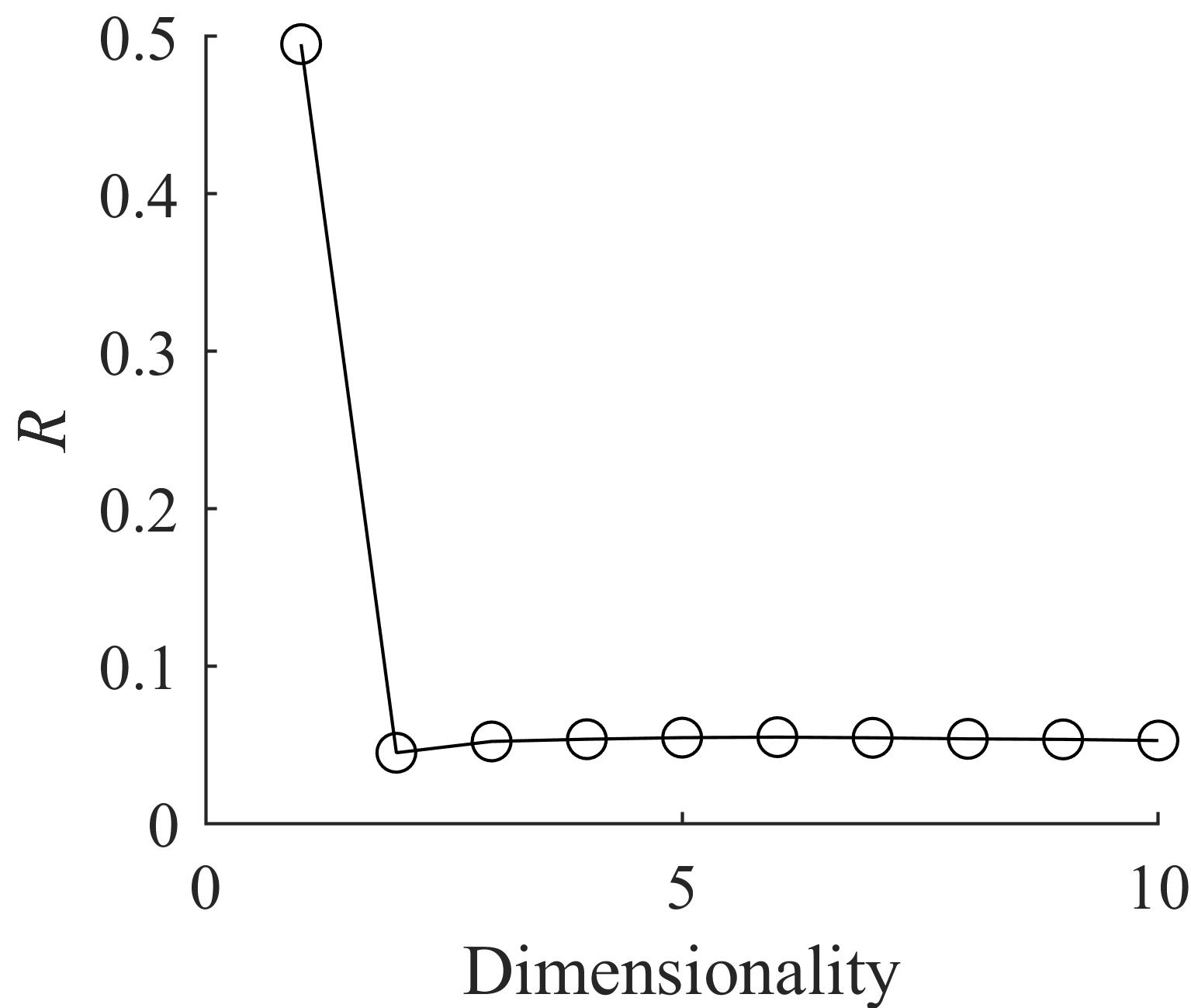}
\caption{Same as figure~\ref{Fig:choiceK}, but for the velocity sensor signal manifold.}
\label{Fig:RSensor}  
\end{figure}

In this section, we present the modeling results for the flow field data in \S~\ref{sec4.1}. 
Here, the training velocity field is first normalized and then represented on a least-order manifold. 
In \S~\ref{sec4.2}, we discuss flow estimation, where the full-state flow field is estimated using sensor signal inputs. 
The sensor signals are initially non-dimensionalized based on the estimated wind speed, enabling them to map with the corresponding latent variables. 
These latent variables are then decoded into full-state data using the manifold model.

\subsection{Flow manifold on the rooftop of a UAV vertiport}
\label{sec4.1}

In this section, we model the training data of flow above a UAV vertiport by the non-dimensionalized manifold model.
The results are achieved using the following parameters: the number of clusters $K_c = 100$, the number of nearest neighbors for the encoder $K_e = 2$, and the number of nearest neighbors for the full manifold projection $K_p = 20$. 
In our previous studies~\citep{farzamnik2023snapshots}, any value of $K_e$ within a valid range results in a low residual variance.
Here, we select the lower bound of this range, defined as the smallest $K_e$ that ensures the neighboring graph $\bm{G}$ is connected. The choices of $K_c$ and $K_p$ serve to balance computational load and accuracy.
The data are first non-dimensionalized by the corresponding oncoming velocity $U_\infty$ and then put into the manifold learner.
Figure~\ref{Fig:choiceK} illustrates the residual variances for different ISOMAP dimensions derived from the coarse-grained centroid manifold. 
For this configuration, we identify the optimal dimension $p = 2$ with $R = 0.041$, reflecting the true dimensionality of the problem. 
The residual variance remains nearly constant for $p > 2$, suggesting that the input data manifold, despite being embedded in a much higher-dimensional space, effectively has only two significant dimensions.
Consequently, the underlying structure of the data can be well represented using a least-order approach, simplifying the analysis without substantial loss of information.
Notably, applying POD to this dataset also reveals two dominant modes, typically with over $99\%$ of the fluctuation energy captured. 
However, the number of leading POD modes does not always align with manifold dimensionality, highlighting the differences in the information each approach captures. 
For a deeper exploration of this topic, see reference~\citep{marra2024actuation}, where the relationship between POD modes and manifold dimensions is explored extensively.

Figure~\ref{Fig:manifolds} illustrates the phase portrait of the coarse-grained centroid manifold and the projected full manifold. 
Here, $\gamma_1$ and $\gamma_2$ represent the $2$ latent variable coordinates. 
It is worth noting that the values of $\gamma$ are determined by the norm between datasets, which varies based on configurations and has no specific physical interpretations. 
Thus, the focus should be on their relative values and relationships between different variables.
Both manifolds exhibit a circular structure, indicating symmetry and clear periodicity within the dataset. 
This circular shape suggests that the underlying dynamics of the system are regular and predictable, consistent with the periodic wind angle information contained in the data. 
The snapshot manifold preserves the general circular shape of the centroid manifold while incorporating finer details, resulting in a smoother manifold shape.
These additional details may capture subtle variations within the data, offering a refined view of the system's behavior.
However, the manifold is not entirely uniform or continuous. 
This irregularity stems from the geometric configuration of the vertiport, which leads to a non-uniform distribution of centroids that affects the shape and continuity of the full snapshot manifold. 
Consequently, certain areas may appear more compressed or stretched.

Investigating whether the manifold coordinates correlate with relevant flow quantities is crucial. Figure~\ref{Fig:manifoldsWindConditions} shows the relationship between the latent variable $\bm{\gamma}$ and wind conditions for the full manifold, with data points color-coded for different wind speeds and angles. 
Figure~\ref{Fig:manifoldsWindConditions} (a) indicates that oncoming wind speed has little effect on both coordinates $\gamma_1$ and $\gamma_2$, as evidenced by the significant overlap of data points across varying wind speeds. 
This suggests independence between oncoming wind speed and the latent variables, supporting the model's effectiveness in incorporating and validating Reynolds number independence.

Conversely, a strong correlation appears between the wind angle and the latent variables, as expected. 
The circular and symmetric shape of the manifold mirrors the nature of wind angles, indicating that the complex, high-dimensional velocity field around the vertiport can be reduced to a least-order problem primarily driven by wind angle. 
This simplification enhances our understanding and modeling of similar configurations, like the building cluster in Appendix~\ref{App1}, and further leads to more efficient analysis and predictions. 

\begin{figure}
\includegraphics[width=8cm]{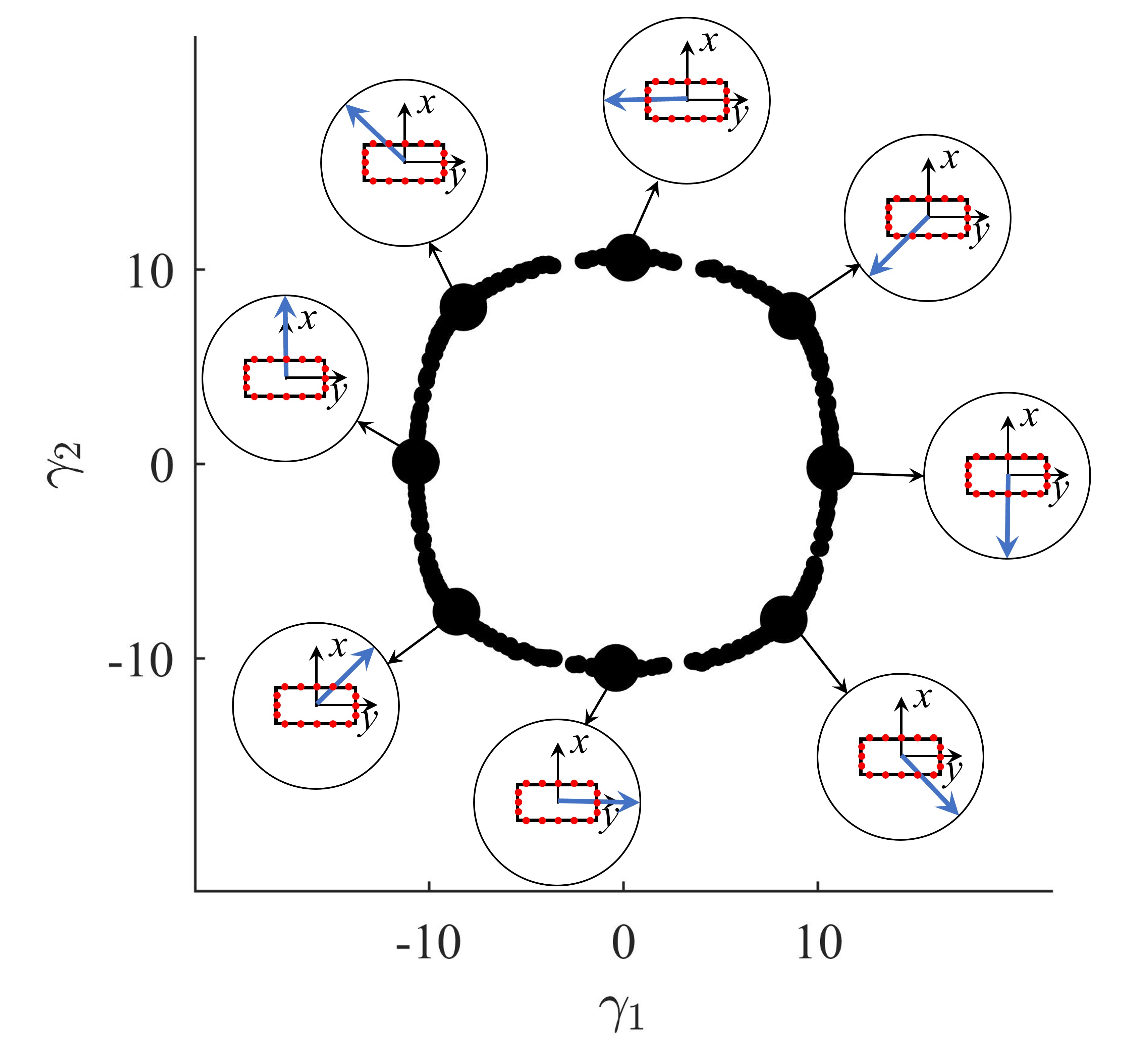}
\caption{Manifold of the velocity sensor signals, with the oncoming wind angles of the large dots plotted around the manifold.
The manifold shape is similar to the snapshot data manifold, suggesting that the flow characteristics are captured.
}
\label{Fig:manifoldsSensor}  
\end{figure}
The sensor signals are also taken for modeling. This step helps identify data patterns from the sensors and verify their validity for flow estimation, though it is not essential to the overall framework. 
Due to the smaller size of the sensor dataset, the snapshot manifold can be directly obtained with $K_e = 20$.
Figure~\ref{Fig:RSensor} displays the residual variance of the velocity sensor, revealing a similar $2$-D manifold as seen in the velocity field data. 
This similarity indicates that the flow characteristics are accurately captured, confirming the effectiveness of the current sensor placement and the comprehensiveness of the sensor signals. 
As shown in the sensor signal manifolds in Figure~\ref{Fig:manifoldsSensor}, the shape remains circular and symmetric. The overall shape appears slightly square, reflecting the rectangular distribution of the sensor locations around the vertiport.
Additionally, the circular shape of the manifold allows for the extrapolation of higher wind speeds through non-dimensionalization, enabling the estimation of rare wind conditions. 
This capability to predict under unusual scenarios, such as wind gusts, highlights the robustness of the manifold-based approach.

\subsection{Flow estimation on the rooftop of a UAV vertiport from sensor signals}
\label{sec4.2} 
In this section, the non-dimensionalized manifold is employed for flow estimation from sensor signals. 
For a given testing sensor signal, the process begins by estimating the oncoming wind speed using the wind condition estimator. 
Next, the corresponding latent variables are obtained from the non-dimensionalized sensor signal. 
These latent variables are then decoded into the full-state velocity field using the manifold learner. 
Here, the number of nearest neighbors for the decoder is set at \( K_d = K_p = 20 \).

The estimation error of the velocity fields (representation error) is evaluated using the normalized mean squared error (NMSE), defined as:
\begin{equation}
    E=\frac{1}{|T|} \sum_{u (t) \in T} \frac{\|u (t)-\hat{u} (t)\|_{\Omega}^{2}}{\|u (t)\|_{\Omega}^{2}} 
\end{equation}
where $T$ is a test set of observations (velocity fields), and $|T|$ is the cardinality of $T$.

\begin{figure*}
\includegraphics[width=15cm]{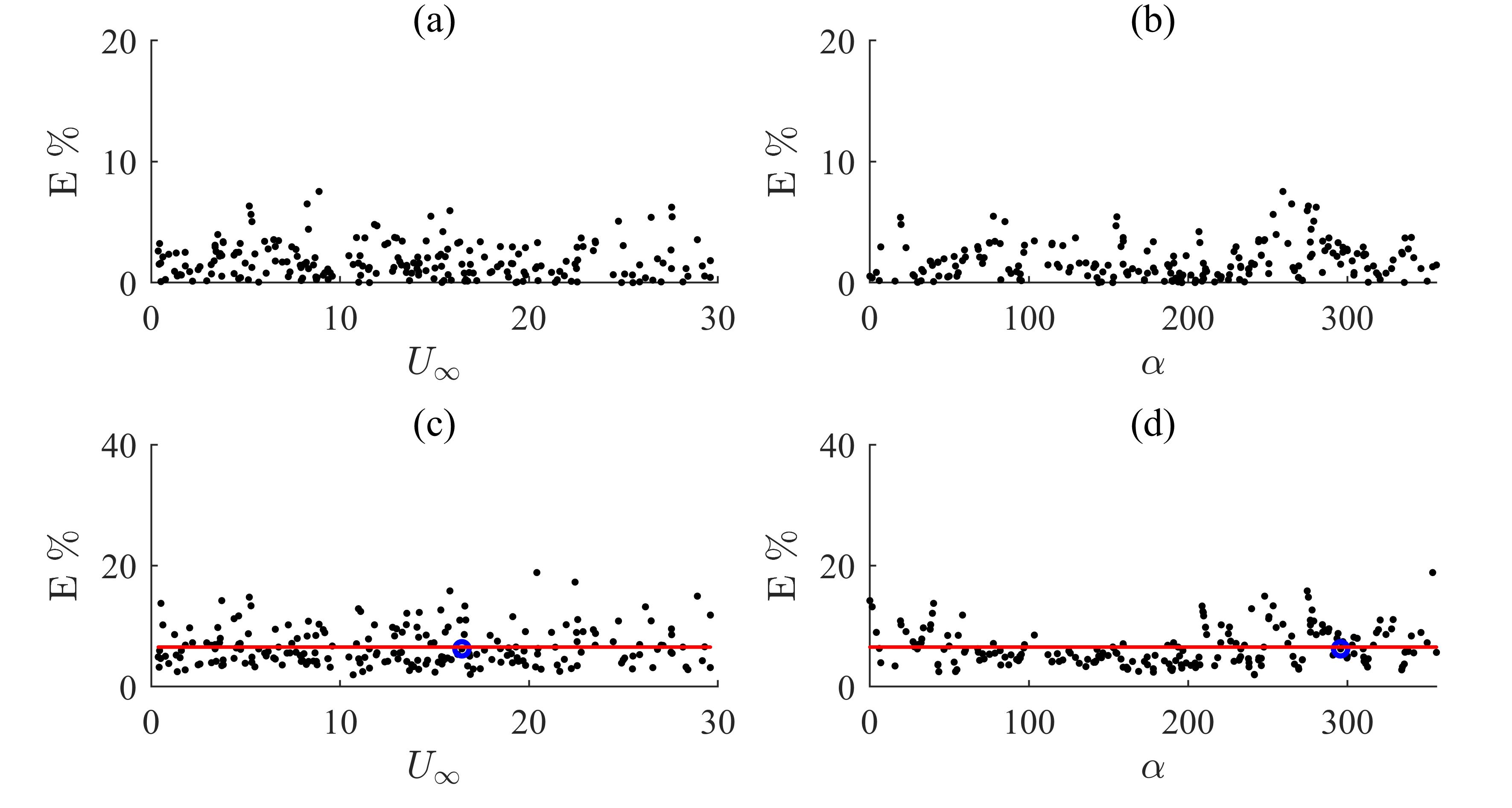}
\caption{Estimation error of test dataset 4:
The estimation error of the oncoming velocity $\hat{U}_\infty$ is shown in (a) for wind speed $U_\infty$ and in (b) for wind angle $\alpha$. 
The normalized squared error (representation error) of the velocity field is presented in (c) against wind speed $U_\infty$ and in (d) against wind angle $\alpha$. 
The normalized mean squared error is highlighted in red ($E = 6.55\%$), while the blue dot indicates the snapshot for further visualization.}
\label{Fig:ErV}  
\end{figure*}
\begin{figure*}
\includegraphics[width=0.9\textwidth]{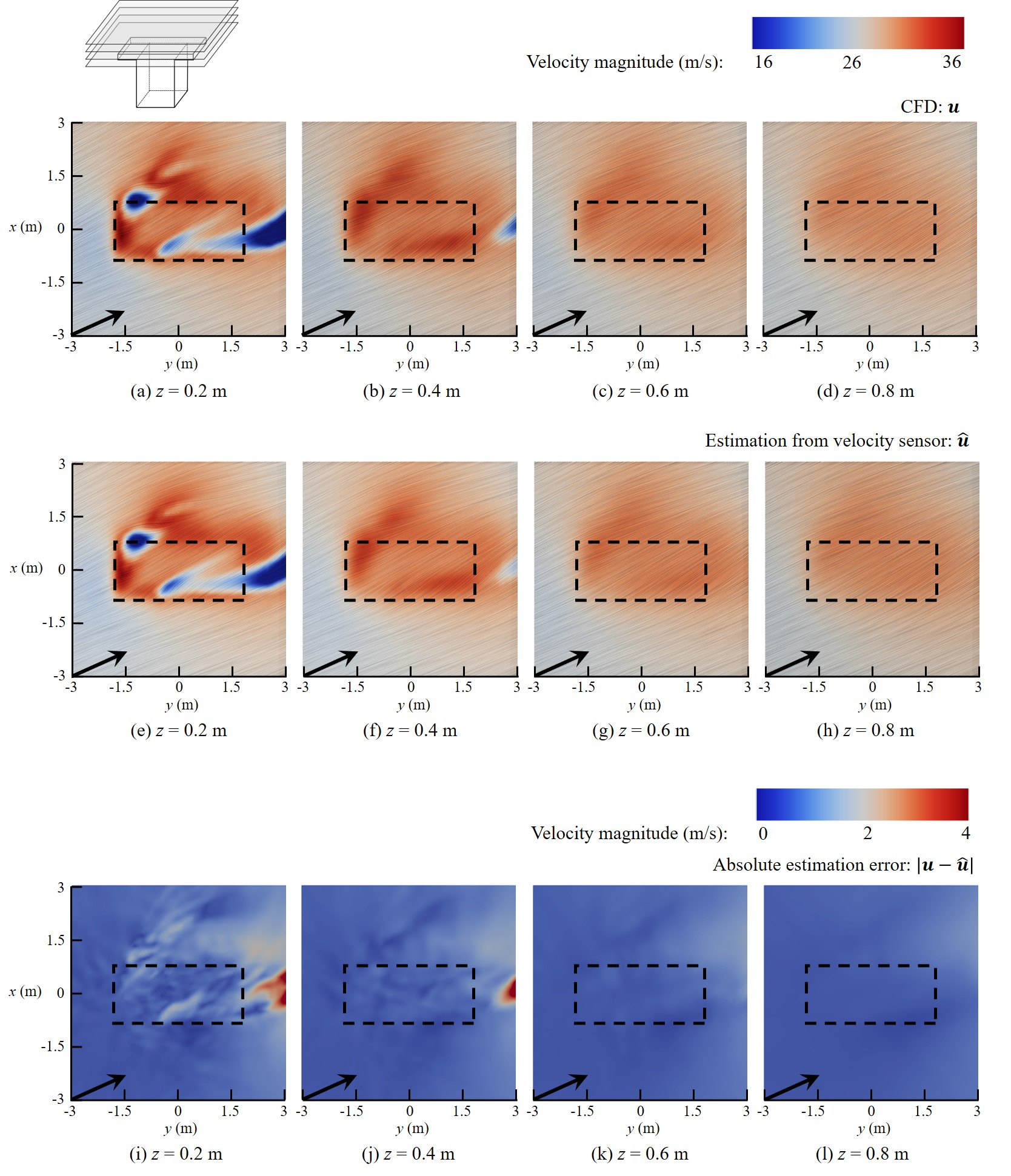}
\caption{Estimated velocity field using the velocity sensor.  
The top left shows the slice locations. The dashed line in the flow visualizations indicates the position of the vertiport.
The arrow starting from the bottom left point on the visualizations indicates the wind angle.
Top: CFD dataset.  
Middle: estimated field.
Bottom: the absolute estimation error.
Snapshot wind conditions: $U_\infty = 26.55$~m/s, $\alpha = 65.22^\circ$, with an estimation error of $E = 2.25\%$.
}
\label{Fig:EstimationV}  
\end{figure*}
Figure~\ref{Fig:ErV} shows the estimation error of $\hat{U}_\infty$ and $\hat{u}$ for all test wind conditions in testing dataset 4 (data for extrapolation).
The error distribution is relatively uniform across various wind speeds and angles, indicating no specific conditions with significantly higher errors, which demonstrates the generalization capability of the framework.
The most significant error among all estimated velocity fields is within $10\%$, while most have errors below $5\%$.
Even for the extrapolated wind conditions which are not included in the CFD database, the error distribution remains consistent with the rest of the dataset.
An estimated snapshot, marked in blue and close to the normalized mean squared error, is selected for flow visualization in Figure~\ref{Fig:EstimationV}. 
The absolute estimation error $\left | \bm{u}-\hat{\bm{u}} \right |$ is plotted in the bottom row.
This visualization includes the estimated flow field across four rooftop transverse planes at heights of $z = 0.2$~m, $0.4$~m, $0.6$~m, and $0.8$~m. 
The accurate prediction of the wind angle is evident, with the flow structure in the estimated data closely resembling that of the CFD data, highlighting the framework's effectiveness in capturing full-state flow dynamics.

\subsection{Error sources of the model}
\label{sec4.3}

\begin{table*}[ht]
\caption{Estimation error of the oncoming wind speed and velocity field by different cases from the velocity sensor. For details, see text.}
\label{Table4}

\resizebox{\textwidth}{!}{
\begin{ruledtabular}
\begin{tabular}{l|cccc|c|c}
\textbf{Case} & \textbf{Oncoming velocity}
              & \textbf{Manifold type}
              & \textbf{Data type}
              & \textbf{Estimation type}
              & $\hat{U}_\infty$ \textbf{Error} (\%)
              & \textbf{Error} (\%)   \\[2pt]
\hline
Case 1 & Single
       & Full
       & Training
       & Interpolation
       & 0
       & 0          \\[2pt]
\hline
Case 2 & Multiple 
       & Full
       & Training
       & Interpolation
       & 0
       & 0          \\[2pt]
\hline
Case 3 & Multiple
       & Coarse-grained
       & Training
       & Interpolation
       & 0
       & 0.551       \\[2pt]
\hline
Case 4 & Multiple
       & Coarse-grained
       & Testing
       & Interpolation
       & 0.643
       & 1.922       \\[2pt]
\hline
Case 5 & Multiple
       & Coarse-grained
       & Testing
       & Extrapolation
       & 0.650
       & 1.972      \\[2pt]
\end{tabular}
\end{ruledtabular}
}
\end{table*}
To gain a comprehensive understanding of the framework, we compare $5$ cases with different model structures and data categories. 
Table~\ref{Table4} displays the mean estimation error for the various combinations within the framework. 
The results are obtained for the velocity sensor; for the pressure sensor, see Appendix~\ref{App2}.
Here, the full manifold refers to using all velocity field data $\bm{u}^{*m}$ in the modeling process (standard ISOMAP), while the coarse-grained manifold indicates using the cluster centroids $\bm{c}^*_k$ (cISOMAP). 
The training data utilizes the existing database, with single $U_\infty$ set at $b_4 = 7.9$~m/s and multiple $U_\infty$ selected randomly. 
The testing data is gathered from additional simulations under random wind conditions, with interpolation covering wind speeds from $b_0$ (0.2 m/s) to $b_8$ (20.7 m/s) and extrapolation from $U_\infty = 20.7$~m/s to $U_\infty = 30$~m/s. 
For the training data, the nearest neighbor for the wind condition estimator is $K_w = 1$, while for the testing data, $K_w = 2$. 
The comparison between the full manifold and the coarse-grained manifold evaluates the impact of the clustering approach, while comparisons across different data sources reveal training and generalization errors.
Additionally, comparing non-dimensionalized multiple wind speeds with a single wind speed validates Reynolds number independence, and extrapolation tests performance under rare wind conditions.
Comparing Case $1$ and Case $2$, model accuracy with multiple non-dimensionalized wind speeds is the same as that with a single speed (all $0$ \% error), indicating a low training error from the full manifold.
This independence simplifies the manifold structure and ensures consistent performance across varying wind speeds.
The comparison of Case $2$ and Case $3$ shows a $0.55 \%$ increase in error with coarse-graining.
Despite this increase, clustering typically reduces computational load by one order of magnitude based on the number of clusters. 
This substantial computational efficiency makes cISOMAP particularly valuable for handling large-scale datasets, emphasizing its practical benefits in industrial applications, where computational power and time are critical constraints.
Comparing Case $3$ and Case $4$ shows that the generalization error remains within acceptable limits. 
The observed error primarily arises from biases introduced by the wind condition estimator rather than from the manifold model itself. 
The increase in error also indicates that improvements in the estimator could further enhance overall accuracy.
Finally, the comparison between Case $4$ and Case $5$ demonstrates that the framework is highly resilient to extrapolation, which also validates Reynolds number independence. 
The minimal additional estimation error observed when extrapolating to rare wind conditions highlights the robustness of the model, ensuring reliable predictions even under less common scenarios. 
This resilience reveals the practical value of the framework in real-world applications where encountering extreme conditions is essential for further operations.

\section{Conclusions}
\label{sec5}
In this study, we propose a data-driven framework that extrapolates full-state velocity fields from limited sensor signals using a non-dimensionalized ISOMAP-$k$NN manifold learner. 
A key challenge in data-centric approaches is that rare events, such as high or low velocities, are not adequately represented in the database.
We address this challenge by exploiting that non-dimensionalized turbulent flows are nearly independent of the  Reynolds number. 
Thus, wind conditions far beyond the available training data can be extrapolated.

The proposed framework approximates the mean flow data from many operating conditions as a non-dimensionalized coarse-grained manifold.
The first encoder normalizes the flow data with the oncoming wind speed.
The second encoder projects this non-dimensionalized data into a least-order manifold via ISOMAP.
This manifold is coarse-grained with cluster-based collocation points to reduce the computational load of the en- and decoding.
The decoders reverse this process: $k$NN performs the initial decoding, followed by a second decoder that restores the normalization.
This manifold simplifies the flow estimation
from sensor signals.
First, these signals are employed to estimate the oncoming wind speed.
We emphasize that the assumed Reynolds-number independence allows us to infer wind speeds far beyond the database.
Second, 
the correspondingly non-dimensionalized sensor signals infer the latent variables on the manifold. 
These latent variables are decoded into full-state flow fields as described above. 
Reynolds number independence enables this methodology to extrapolate wind conditions outside the training set, a feature that traditional data-centric frameworks cannot achieve.

We demonstrate the effectiveness of this framework for the flow on the rooftop of a vertiport
at Reynolds number from $4 \times 10^4$ to $6 \times 10^6$.
Flows with 
1080
 different oncoming velocities and angles  are simulated
with Reynolds averaged Navier-Stokes simulations for the same inlet boundary layer.
Hence, the flow is expected to lie
on a two-dimensional manifold with two latent variables 
that exhibit a strong correlation with the angle
of the oncoming flow.
Intriguingly, 
the non-dimensionalized manifold learner
corroborates Reynolds number independence:
a limit-cycle manifold 
leads to a representation error
of about $2\%$, 
i.e.,\ the non-dimensionalized flow is effectively a function of wind direction.

The error analysis indicates that the framework enables accurate extrapolation, maintaining performance even in wind conditions far away from the database.
This feature significantly reduces
the amount of testing data for the estimator
while improving estimation accuracy.
Accurate wind gust prediction 
is critical for the safe operation of drones above vertiports for wind levels around $b_6$ and larger. 
The landing drone experiences a flow velocity that reverses in the recirculation zone.
This can lead to a large displacement error 
for gust-agnostic flight control.
To our knowledge, this is the first demonstration of 
an extrapolatory flow estimation that exploits a non-dimensionalized manifold and 
the first flow estimation was conducted on the rooftop of a vertiport.

Summarizing, the proposed estimation framework provides a robust, interpretable, and generalizable approach to flow estimation and modeling. 
The results not only enhance our understanding of urban wind fields but also open avenues 
for applications such as gust-safe drone delivery and flight control. 
Future work  aims to extend this estimation framework 
to predict the wind for drone trajectories in Shenzhen.
These efforts include multiple building clusters, 
 variable wind profiles
 and flow instabilities.

\appendix

\section{Modeling of flow around a cluster of buildings.}
\label{App1}

\begin{figure}[h]
\includegraphics[width=8.5cm]{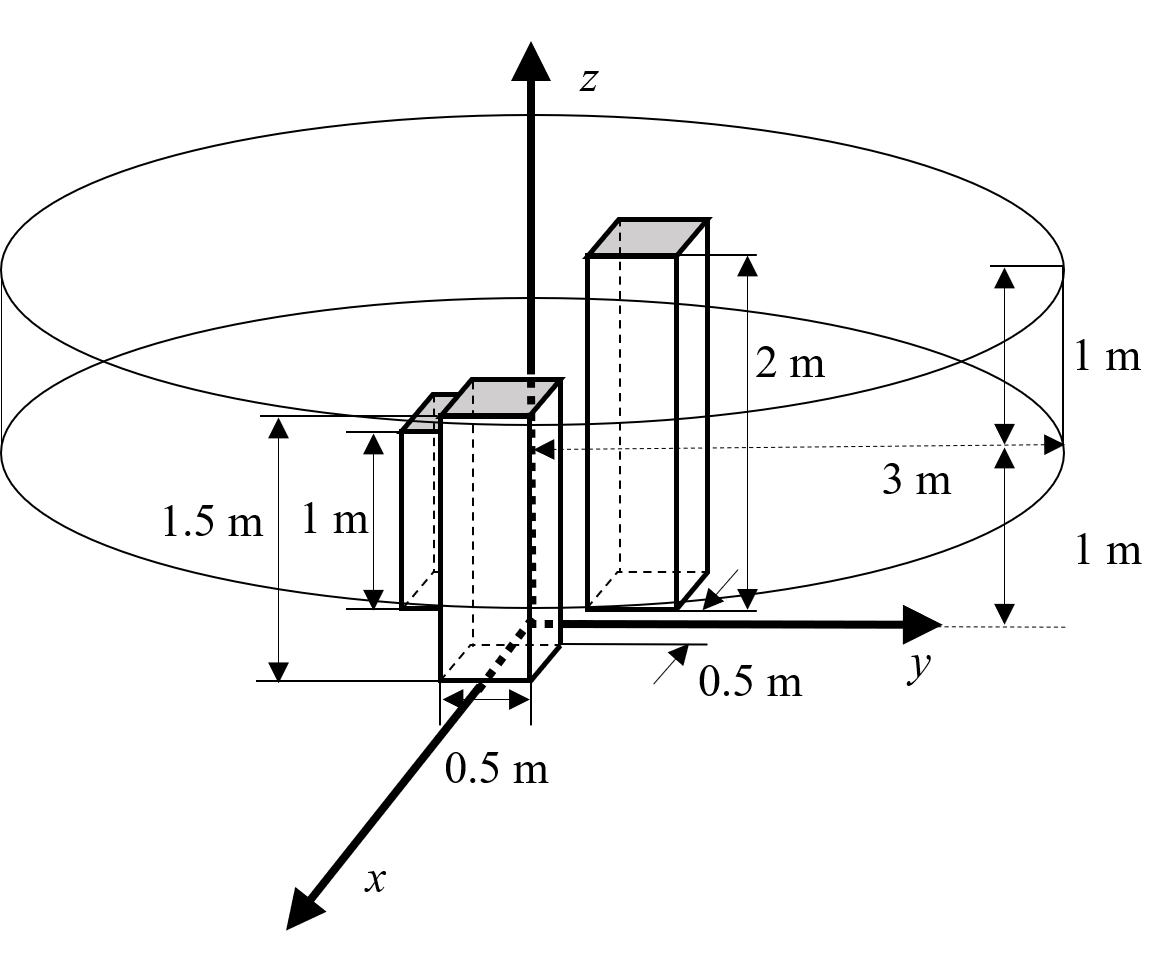}
\caption{The scaled building cluster configuration:
The three buildings are arranged in an equilateral triangle: the front building is $1.5$~m high, the left rear building is $1$~m high, and the right rear building is $2$~m high. 
Flow field data are collected from a cylindrical region with a radius of $3$~m and a height of $1$~m, extending from the rooftop of the right rear building to that of the left rear building.
}
\label{Fig:building}  
\end{figure}

Figure~\ref{Fig:building} shows the configuration of the building cluster.
The configuration comprises $3$ buildings distributed at the three vertices of an equilateral triangle.
The buildings are $1$~m, $1.5$~m, and $2$~m high.
The field data is set as a cylinder with a radius of $3$~m and a height of $1$~m.
The bottom of the cylinder is set on the rooftop of the 1~m-high building.
The dataset includes snapshots for 4 oncoming velocities ($10.7$~m/s, $13.8$~m/s, $17.1$~m/s, and $20.7$~m/s) and $120$ incidence angles (incremented by $3^{\circ}$), totaling $480$ flow fields.

\begin{figure}[h]
\includegraphics[width=6.5cm]{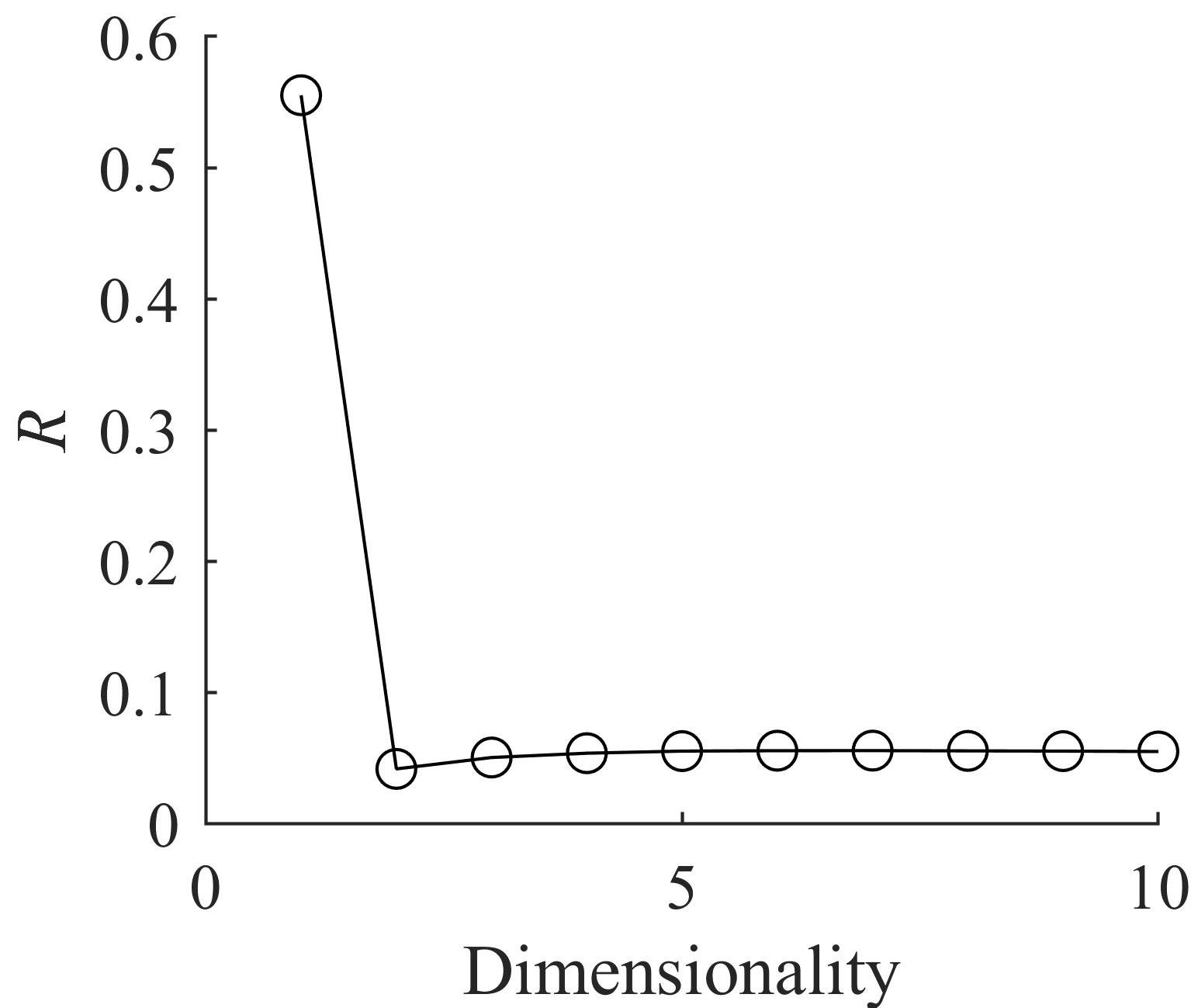}
\caption{Same as figue~\ref{Fig:choiceK}, but for the building cluster.}
\label{Fig:choiceK_building}  
\end{figure}

\begin{figure}[h]
\includegraphics[width=6.5cm]{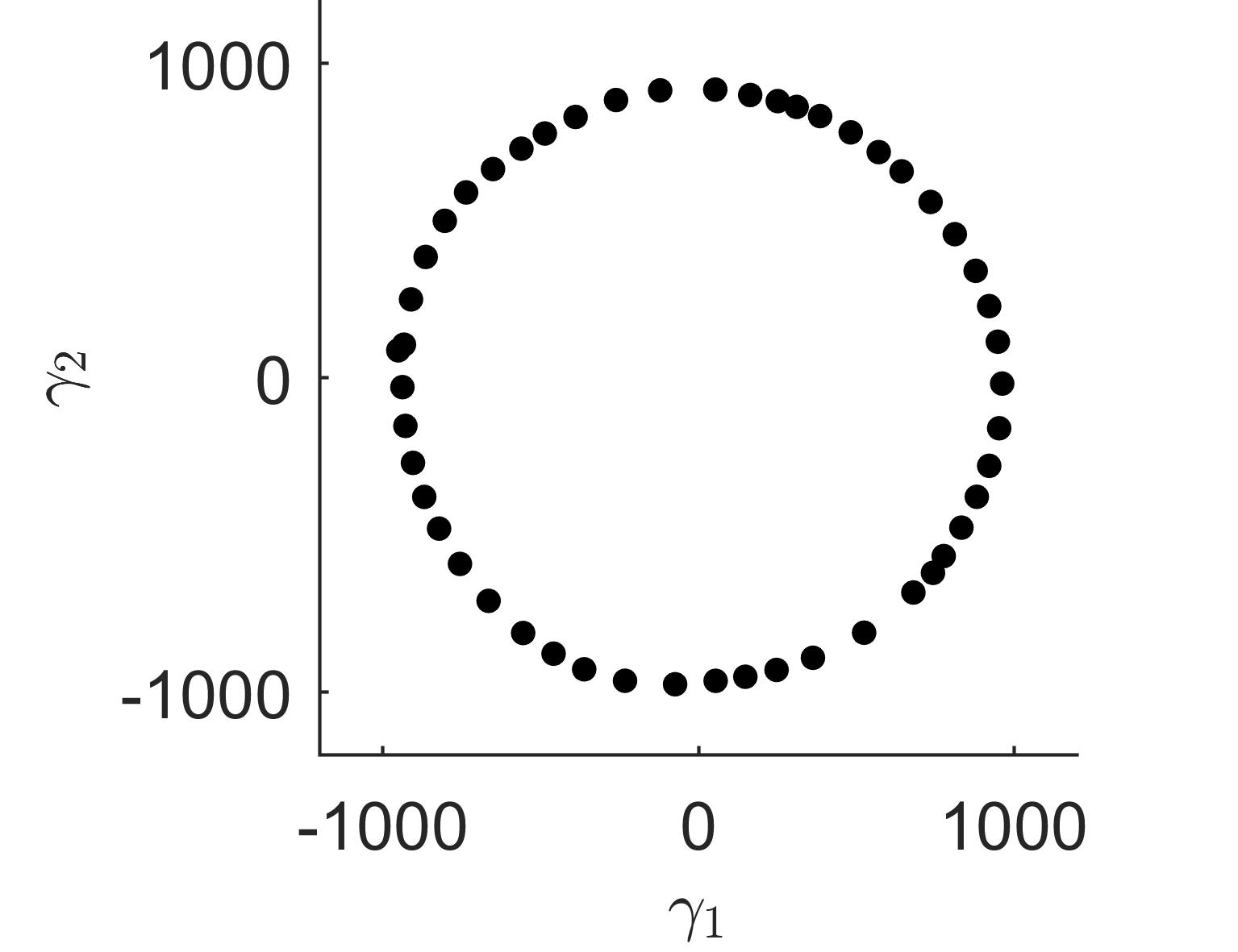}
\includegraphics[width=6.5cm]{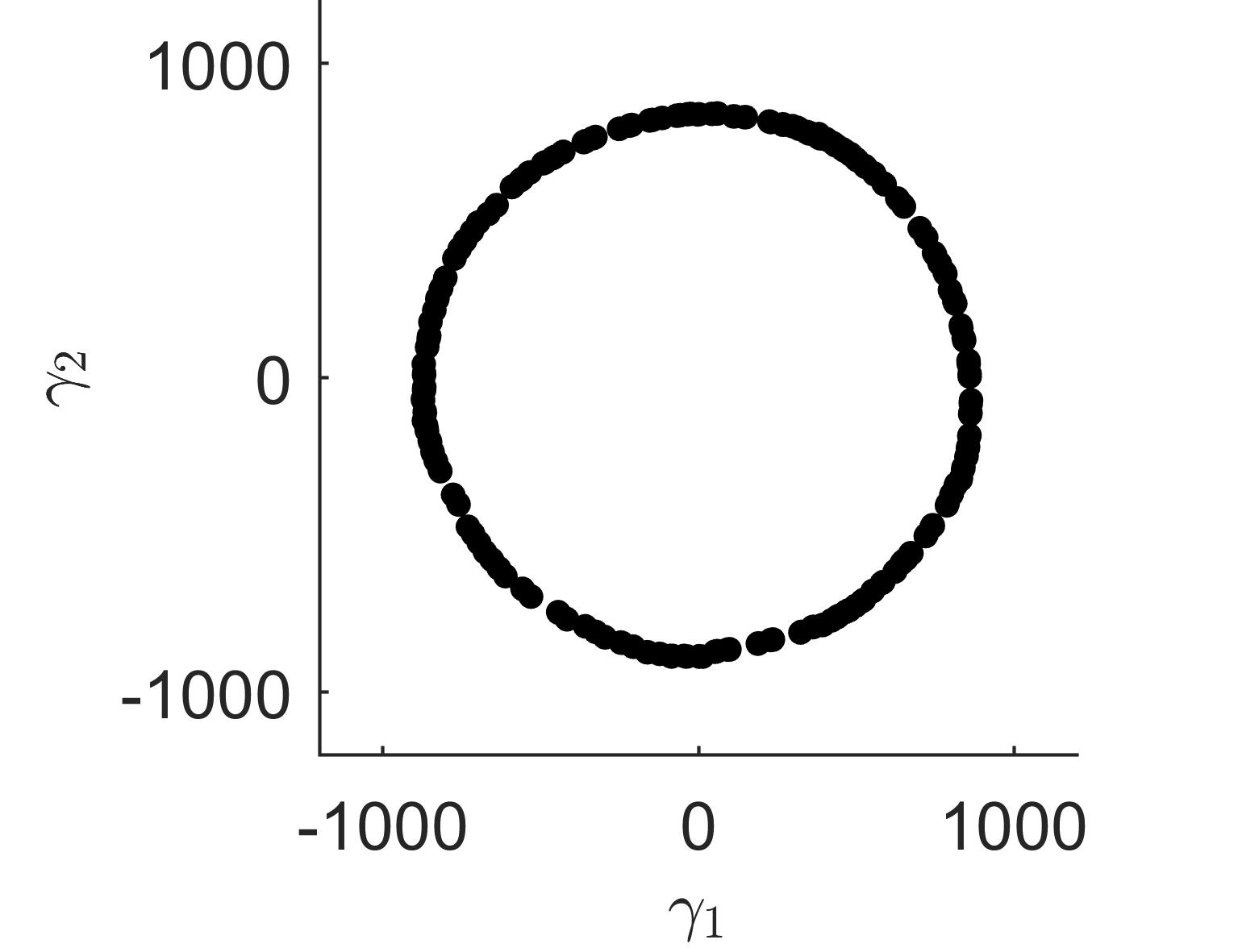}
\caption{Same as figure~\ref{Fig:manifolds}, but for the building cluster.}
\label{Fig:manifolds_building}  
\end{figure}

\begin{figure}[h]
\includegraphics[width=6.5cm]{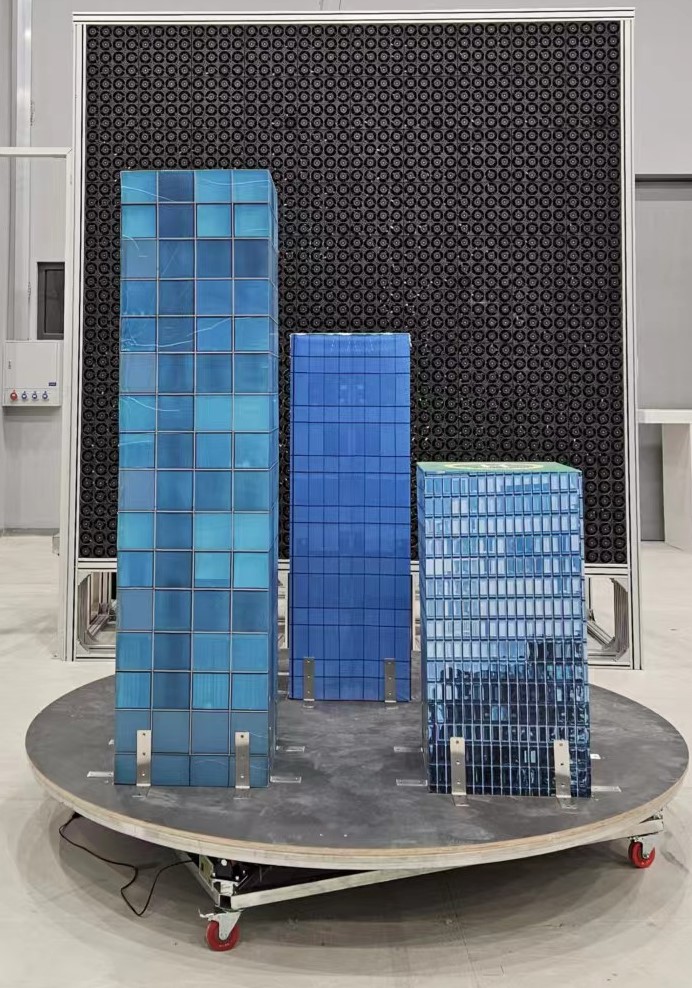}
\caption{Experimental configuration of the scaled building cluster configuration on a rotating `round dish'.
}
\label{Fig:buildingPhoto}  
\end{figure}

Figure~\ref{Fig:choiceK_building} shows the reduced dimension and the residual variance and figure~\ref{Fig:manifolds_building} shows the snapshot manifold and the sensor manifold of the building cluster.
The result is computed with $K_c = 60$, $K_e = 2$ and $K_d = 2$.
Based on the same least-order manifold, it can be inferred that the current framework is adaptable to similar configurations. 
This versatility suggests that the framework can be extended to applications such as urban wind prediction, where understanding and modeling wind patterns in complex environments is crucial.
In our future work, we schedule to conduct related experimental studies with the rotating building cluster configuration and the fan-array wind generator~\citep{liu2024aerodynamic}, as illustrated in Figure~\ref{Fig:buildingPhoto}.

\section{Flow estimation by the pressure sensor}
\label{App2}
For the pressure sensor, the data scaling \eqref{eq:Vscaling} becomes
\begin{equation}
s^* = s/ (\rho U_\infty^2),
\label{eq:Pscaling}
\end{equation}
and the scaling factor $\lambda^l$ in the wind condition estimator is also correspondingly changed following:
\begin{eqnarray}
\lambda^m (t) = \arg \underset{\mu }{\min} \left\| \mu^2 \bm{s}^{m} - \bm{s}(t) \right\|_2,\\
l(t) = \arg \underset{m}{\min} \left\| {(\lambda^m (t))}^2 \bm{s}^{m} - \bm{s}(t) \right\|_2. 
\end{eqnarray}

\begin{figure}[h]
\includegraphics[width=6.5cm]{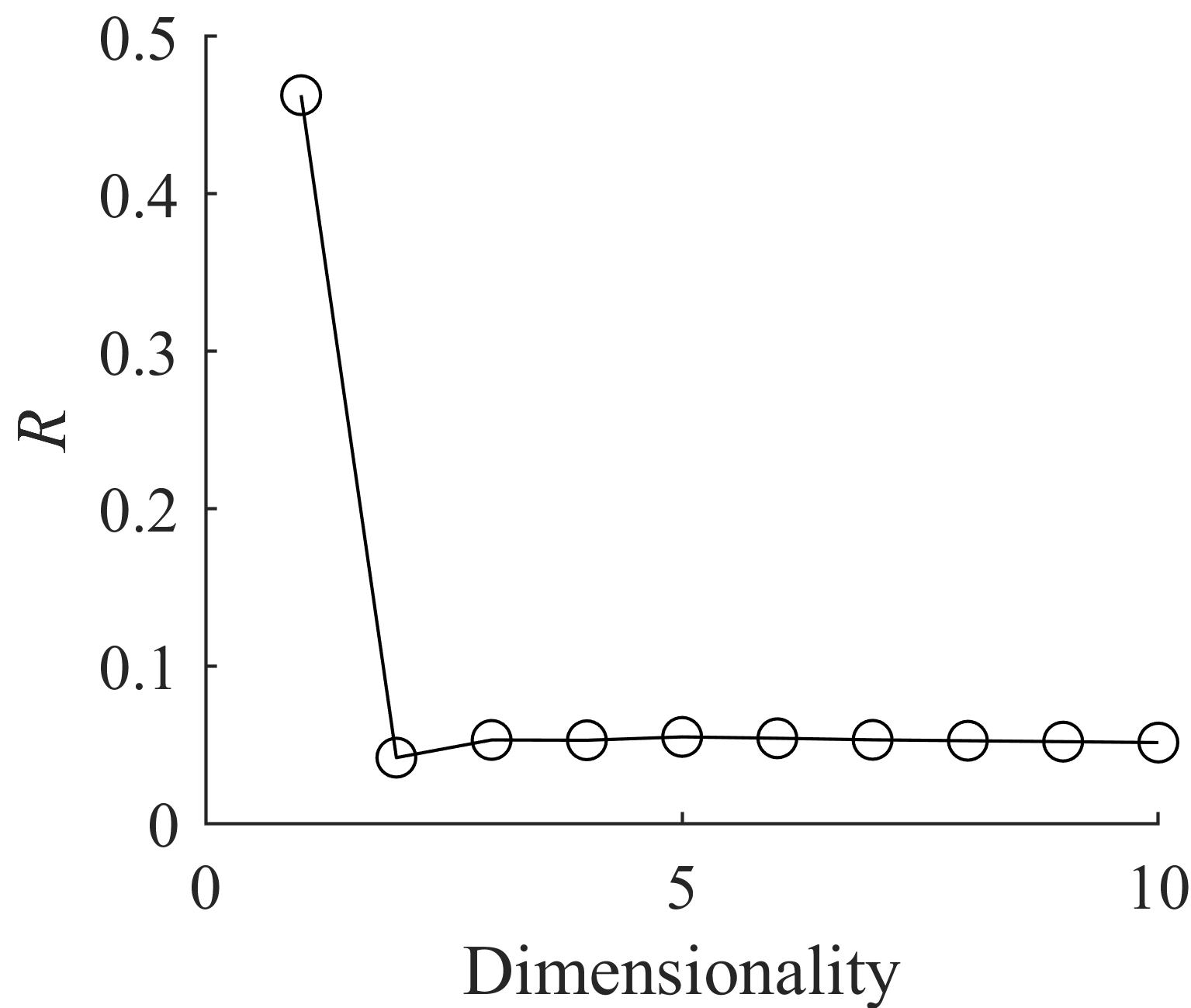}
\includegraphics[width=6.5cm]{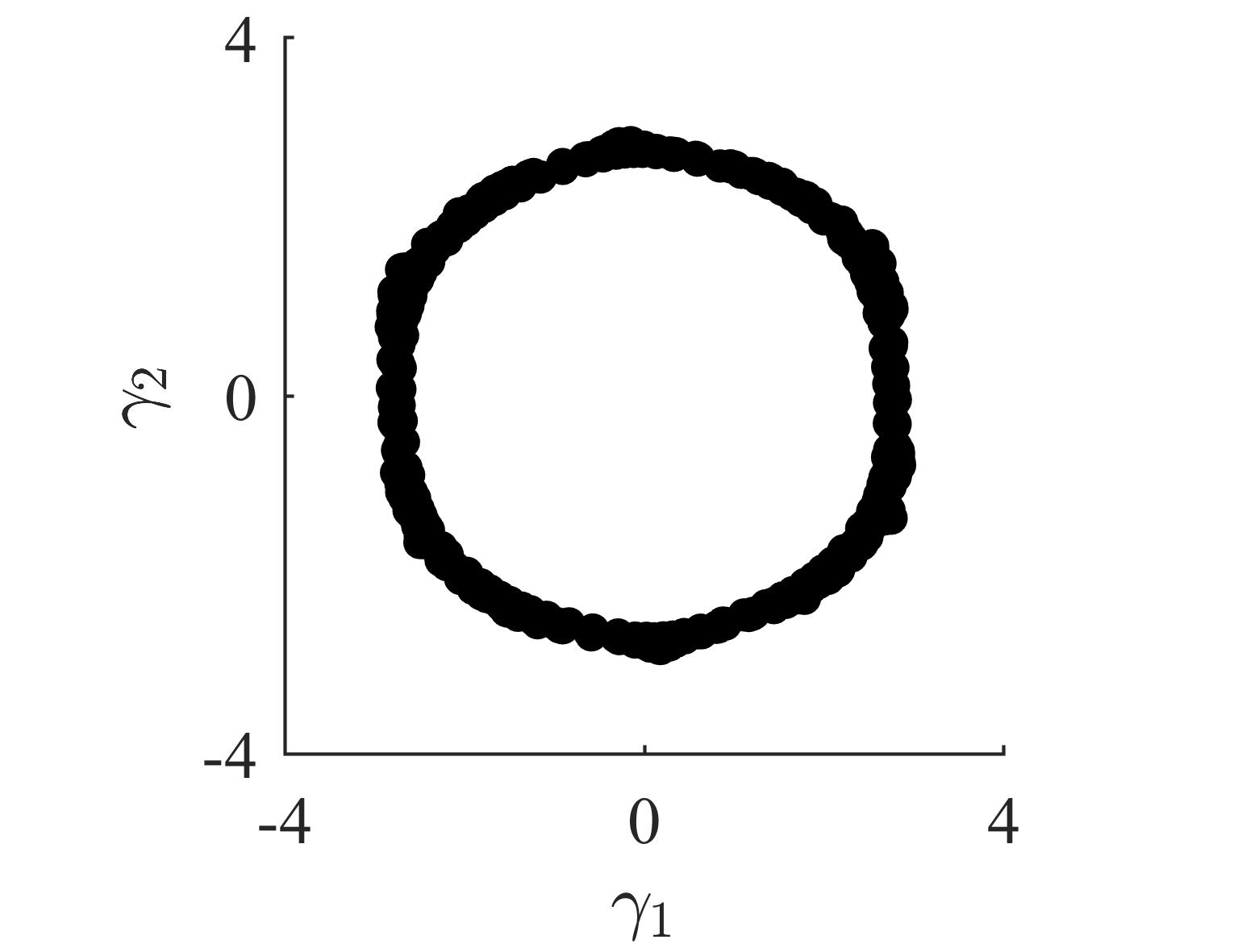}
\caption{Residual variance and the manifold of the pressure sensor signal.}
\label{Fig:PsensorManifold}  
\end{figure}

\begin{figure}[h]
\includegraphics[width=8.5cm]{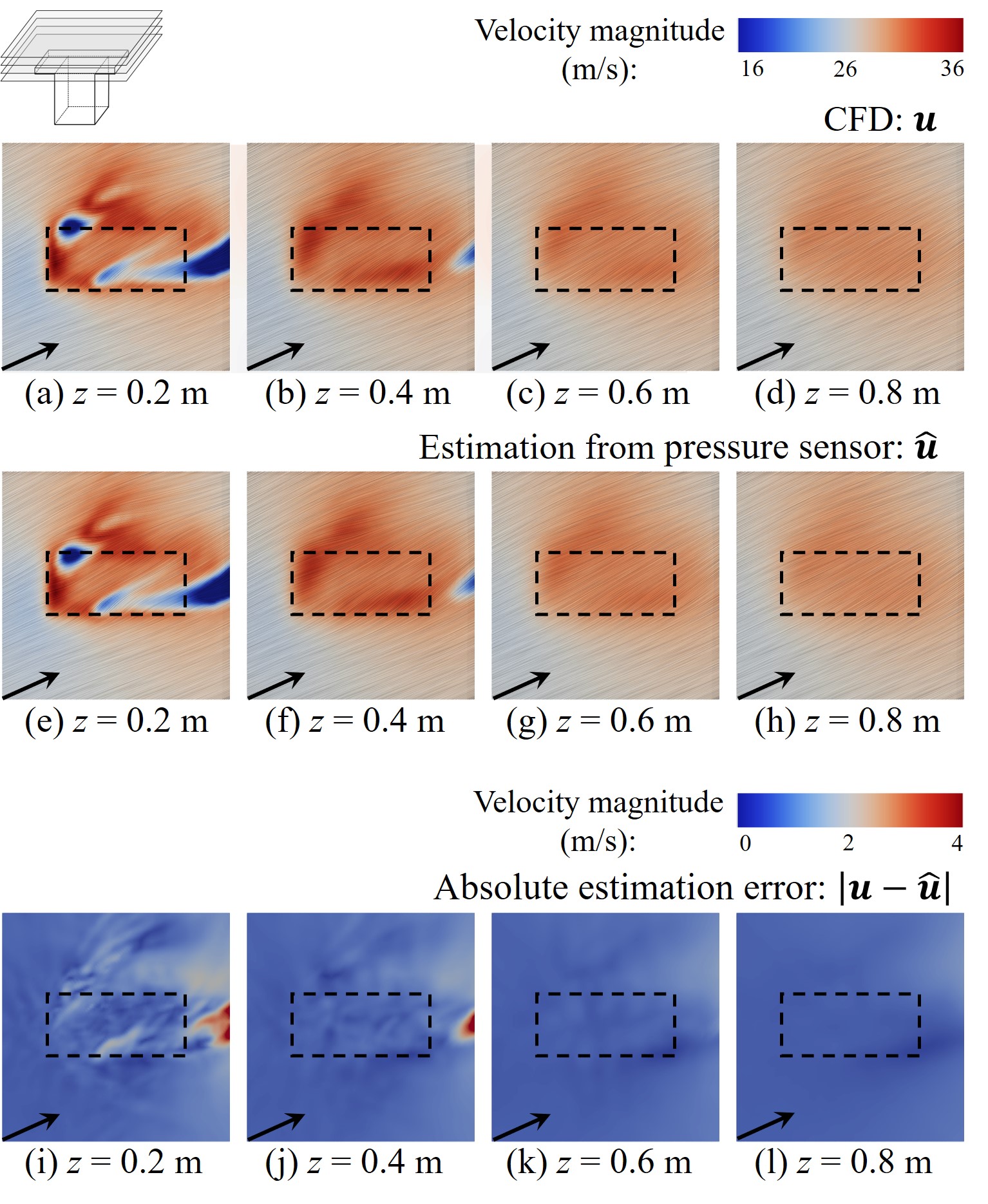}
\caption{The same description of figure~\ref{Fig:EstimationV} applies, but from the pressure sensor, with $E = 2.35\%$.}
\label{Fig:EstimationP}  
\end{figure}

Figure~\ref{Fig:PsensorManifold} shows the residual variance and the manifold of the pressure sensor signal.
Both the dimensions and the shape of the manifold are in high agreement with the results obtained from the velocity sensor, with $K_e = 20$.
For the estimation, table~\ref{Table5} presents the mean errors for various combinations.
\begin{table}[h]
\caption{The estimation error of the oncoming wind speed and velocity field by different cases from the pressure sensor.}
\label{Table5}
\begingroup
\renewcommand{\arraystretch}{1.1} 
\begin{ruledtabular}
\begin{tabular}{lcc}
\textbf{Case}   &  $\hat{U}_\infty$ \textbf{Error} (\%)    & \textbf{Error} (\%) \\[2pt]
\hline
Case 1 & 0 &  0       \\[2pt]
\hline
Case 2 & 0 &  0       \\[2pt]
\hline
Case 3 & 0 &  0.783       \\[2pt]
\hline
Case 4 & 0.654  & 2.082      \\[2pt]
\hline
Case 5 & 0.693  & 2.214      \\[2pt]
\end{tabular}
\end{ruledtabular}
\endgroup
\end{table}
After scaling, the flow field estimation based on the pressure sensor signal achieves the same accuracy as the velocity sensor.
Figure~\ref{Fig:EstimationP} shows the visualization by the pressure sensor from an estimated snapshot with an error close to the normalized mean squared error.
The flow structure is similar to the CFD data,
like the velocity sensor-based reconstruction.
These observations demonstrate the ability of the Reynold-scaled manifold framework to adapt to multiple sensor types.

\section{Robustness analysis}
\label{App3}

\begin{figure}[h]
\includegraphics[width=6.5cm]{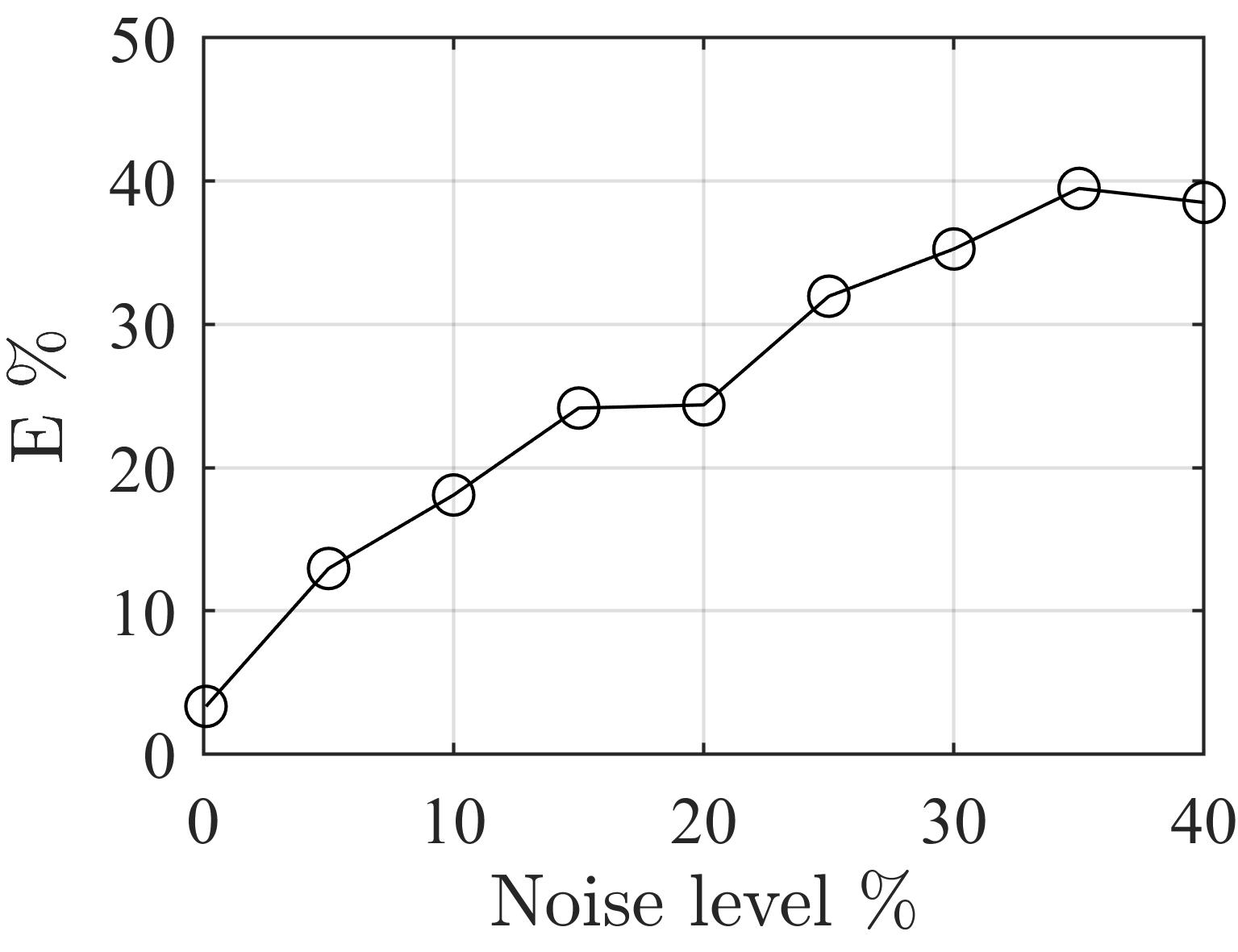}
\caption{Estimation error with noisy input for the velocity sensor in testing dataset 4. 
The noisy sensor signal is created by adding white Gaussian noise of varying levels to the clean signal. 
As a reference, the estimation error of the clean (noise-free) signal is $1.972\%$.}
\label{Fig:EstimationN}  
\end{figure}

In industrial applications, input signals frequently contain noise. 
To assess the robustness of our framework, we superimpose increasing levels of white Gaussian noise, $\bm{n}(t)$, to the clean sensor signal and evaluate its performance. 
The noise level, ranging from $0\%$ to $40\%$, is defined as a proportion of instantaneous signal power:
\begin{equation}
\mathrm{Noise \; level} = P_{\bm{n}}/P_{\bm{s}},
\label{eq:noiselevel}
\end{equation}
where $P_{\bm{n}}$ is the noise power, and $P_{\bm{s}}$ is the clear signal power, for the $t$-th input signal:
\begin{equation}
P_{\bm{n}}(t) = \bm{n}(t)^{\top} \bm{n}(t),
\label{eq:pn}
\end{equation}
\begin{equation}
P_{\bm{s}}(t) = \bm{s}(t)^{\top} \bm{s}(t),
\label{eq:pn}
\end{equation}
The fourth testing dataset with the velocity sensor is used, and the results are shown in figure~\ref{Fig:EstimationN}.
As expected, there is a strong correlation between noise level and estimation error, indicating that the framework is sensitive to noise. 
Interestingly, the relationship is almost linear.
Higher noise levels gradually reduce the model’s accuracy compared to the reference clean data. 
Nonetheless, the model retains acceptable accuracy when noise levels remain below $20\%$.

%

\newpage
\begin{acknowledgments}

This work is supported by the National Science Foundation of China (NSFC) through grants 12172109 and 12302293,
by the Guangdong Basic and Applied Basic Research Foundation under grant 2022A1515011492,
and by the Shenzhen Science and Technology Program under grants JCYJ20220531095605012, KJZD20230923115210021 and 29853MKCJ202300205, by the project EXCALIBUR (Grant No PID2022-138314NB-I00), funded by MCIU/AEI/ 10.13039/501100011033 and by “ERDF A way of making Europe”, and by the funding under "Orden 3789/2022, del Vicepresidente, Consejero de Educación y Universidades, por la que se convocan ayudas para la contratación de personal investigador predoctoral en formación para el año 2022".
Chang Hou appreciates the support from the School of Mechanical Engineering and Automation during his PhD thesis at Harbin Institute of Technology (Shenzhen).
In addition, we appreciate valuable discussions with S.~Lera,  
F.~Raps, D.~Sornette, Y.~Yang, and the CAIA team at HIT. 

The authors declare that there is no conflict of interest regarding the publication of this paper.
\end{acknowledgments}

\section*{Data Availability Statement}
The data that support the findings of this study are available from the first 
author upon reasonable request.

\bibliography{reference}

\end{document}